\documentclass[hyper]{JHEP3}

\usepackage[dvips]{graphicx}
\usepackage{epsfig,multicol,amssymb,amsmath,cancel}

\newcommand\fverb{\setbox\fverbbox=\hbox\bgroup\verb}
\newcommand\fverbdo{\egroup\medskip\noindent%
            \fbox{\unhbox\fverbbox}\ }
\newcommand\fverbit{\egroup\item[\fbox{\unhbox\fverbbox}]}
\newbox\fverbbox

\def\bea{\begin{eqnarray}}
\def\eea{\end{eqnarray}}
\def\beq{\begin{equation}}
\def\eeq{\end{equation}}

\title{Axion model in gauge-mediated supersymmetry breaking and
a solution to the $\mu/B\mu$ problem}

\author{Kwang Sik Jeong and Masahiro Yamaguchi \\
Department of Physics, Tohoku University, Sendai 980-8578, Japan  \\
E-mail :
\email{ksjeong@tuhep.phys.tohoku.ac.jp},
\email{yama@tuhep.phys.tohoku.ac.jp} }

\preprint{TU-879}

\abstract{
We present a simple supersymmetric axion model that can naturally explain
the origin of the Higgs $\mu$ and $B\mu$ terms in gauge mediation while
solving the strong CP problem.
To stabilize the Peccei-Quinn scale, we consider mixing between the messenger
fields that communicate the supersymmetry and Peccei-Quinn symmetry breaking
to the visible sector.
Such mixing leads to the radiative stabilization of the Peccei-Quinn scale.
In the model, a Higgs coupling to the axion superfield generates the $B$
parameter at the soft mass scale while a small $\mu$ term is induced
after the Peccei-Quinn symmetry breaking.
We also explore the phenomenological and cosmological aspects of the model,
which crucially depend on the saxion and axino interactions with the ordinary
particles induced by the Higgs coupling to the axion superfield.
}

\keywords{Supersymmetry breaking, Supersymmetric Standard Model}

\begin{document}

\section{Introduction}

Gauge mediation of supersymmetry (SUSY) breaking
\cite{gauge-mediation-classic,gauge-mediation,review-gauge-mediation} is
an attractive mechanism to generate soft terms in the minimal supersymmetric
Standard Model (MSSM).
In particular, since the transmission occurs through the gauge interactions,
models of gauge mediation solve the SUSY flavor and CP problems.
However, gauge mediation has difficulty in explaining the origin of the Higgs
$\mu$ and $B\mu$ terms, and thus requires an extension.
Since the $\mu$ term breaks the Peccei Quinn (PQ) symmetry \cite{PQ-mechanism},
an interesting possibility is that the presence of a $\mu$ term has the same origin
as the invisible axion \cite{review-axion} solving the strong CP problem
\cite{Kim-Nilles-mechanism,chun-mu-axion,murayama-mu-axion,martin-axion-mu-problem}.
The size of the $B\mu$ term is then determined by how the saxion, the scalar
partner of the axion, is stabilized.
A potential for the saxion is generated only after SUSY breaking because the PQ
symmetry makes the scalar potential flat along the saxion direction in the
supersymmetric limit.
This indicates that it is non-trivial to stabilize the saxion, which is
a gauge singlet, within gauge mediation where soft terms receive contribution
proportional to the gauge couplings.

In this paper, we consider a simple axion model in the framework of gauge mediation
that provides a natural solution both to the $\mu/B\mu$ problem and the strong CP
problem.
The model contains matter fields that communicate SUSY breaking from a hidden sector
to the MSSM, and also those that transmit the PQ symmetry breaking.
They are charged under the Standard Model (SM) gauge groups, and become
massive either by directly coupling to the hidden sector SUSY breaking field
or by coupling to the axion superfield.
The most important property of the model is that there is mixing between these
two classes of messengers.
It is through the mixing that the saxion feels SUSY breaking at the loop level
and acquires soft mass comparable to those of the gauge-charged sparticles.
As a result, the saxion is radiatively stabilized at a scale below or around
the scale of gauge mediation.
In this model, the $\mu$ term arises from an appropriate coupling of
the Higgs doublets to the axion supermultiplet either in the superpotential
or in the K\"ahler potential.
Remarkably, the $B\mu$ term is then generated at the correct mass scale,
thanks to the SUSY breaking in the axion supermultiplet induced by the mixing
between the two messenger sectors.

The MSSM soft terms receive negligible threshold corrections at the PQ
scale, but their renormalization group (RG) evolutions are affected by
the PQ messengers.
In particular, if the saxion has a vacuum expectation value rather close
to the scale of gauge mediation, MSSM scalar masses can receive a sizable
contribution through the hypercharge trace term because the PQ messenger
scalars acquire additional soft masses due to the mixing.
This contribution can make the stau lighter than the gauginos.
In the model, the lightest superparticle (LSP) is given either by the axino,
the fermionic partner of the axion, or by the gravitino depending on the scale
of gauge mediation.
The ordinary sparticles dominantly decay into axinos, not into gravitinos,
through the interactions suppressed by the PQ scale.
Meanwhile, the saxion properties are constrained by various cosmological
considerations.
In case that the Universe is dominated by the saxion, the axion energy
density produced by the saxion decay should be less than that of one neutrino
species to be consistent with the Big Bang nucleosynthesis.
In addition, LSPs from the saxion decay should not overclose the Universe.
To satisfy these constraints, one needs to enhance the saxion coupling to the SM
particles.
This is naturally achieved when the $\mu$ term is generated by a superpotential
interaction between the axion supermultiplet and Higgs doublets.

This paper is organized as follows.
In section 2, we examine how the saxion direction is lifted in the presence
of mixing between the messengers that transmit SUSY breaking
and PQ symmetry breaking to the MSSM sector.
We then show in section 3 that the model, where the PQ scale is radiatively stabilized,
can naturally generate the correct mass scale not only for $\mu$ but also for $B\mu$
in gauge mediation.
Sections 4 and 5 are devoted to the discussion of phenomenological and cosmological
aspects.
We will examine the pattern of sparticle masses, the decay of sparticles into
axinos or gravitinos, and the cosmological constraints on the saxion properties.
An important role is played by the saxion/axino interactions with the MSSM particles
induced by the Higgs coupling to the axion superfield.
The last section is for the conclusion.

\section{Axion in gauge mediation}

To invoke the PQ mechanism within the framework of gauge mediation,
we introduce heavy matter superfields that form vector-like pairs under
the SM gauge groups.
These fields are classified as
\bea
\Phi+\bar\Phi &:& \mbox{SUSY breaking messengers},
\nonumber \\
\Psi+\bar\Psi &:& \mbox{PQ messengers},
\nonumber
\eea
depending on the way of getting massive.
The $\Phi+\bar\Phi$ are vector-like also under $U(1)_{\rm PQ}$ and
directly couple to hidden sector fields that participate in SUSY breaking.
They are the usual messengers for gauge mediation.
On the other hand, the PQ messengers couple to the axion superfield $S$ through
Yukawa interaction and thus acquire heavy mass after PQ symmetry
breaking.

One might think that minimal field content is prepared for the PQ
mechanism\footnote{
To solve the strong CP problem, the PQ symmetry should be anomalous under
QCD interactions.
If $\Psi+\bar\Psi$ are charged under QCD, one obtains a KSVZ-type (hadronic)
axion model \cite{KSVZ}.
A DFSZ-type axion model \cite{DFSZ} is otherwise obtained for the Higgs
bilinear $H_uH_d$ charged under $U(1)_{\rm PQ}$.
}
to work in gauge mediation.
However, previous studies
\cite{saxion-run-away-gauge-mediation,hadronic-axion-gauge-mediation,Chun-axion-mu-gauge-mediation}
have noticed that there should be new interactions
transmitting SUSY breaking to the PQ sector in order to fix the PQ scale\footnote{
See also \cite{axions-in-GM} for other interesting observations on axions in gauge
mediation.
}.
This feature is observed under the assumptions that $U(1)_{\rm PQ}$ is
spontaneously broken by a single field, $S$, and that soft terms only receive
gauge-mediated contribution.
Hence, one may extend the model to include extra SM singlet fields
carrying PQ charge, or add an additional source of SUSY breaking such as
gravity mediation
\cite{hadronic-axion-gauge-mediation,Chun-axion-mu-gauge-mediation}.
Another interesting approach we would like to pursue here is to consider
the case that some messengers of the two sectors, say $\bar\Phi$ and $\bar\Psi$,
have the same charge under all the symmetries of the theory.
Then, there arises mixing between them, which makes $S$ feel
SUSY breaking at the same loop level as SM-charged scalars do through gauge
mediation.
Since SUSY breaking generates a potential for the saxion, such mixing can
play an important role in determining the PQ scale.

In this paper, we consider a simple axion model\footnote{
Though we are assuming that the model has the PQ symmetry to solve the strong
CP problem, it is also possible to consider other cases where $S$ corresponds
to a flaton field driving thermal inflation.
In such case, $U(1)_{\rm PQ}$ needs not be exact, but $U(1)_{\rm PQ}$-breaking
terms should be small enough so that the potential for $S$ can remain approximately
flat in the supersymmetric limit.
Most of our discussion can apply to these flaton models \cite{thermal-inflation}.
The property of the angular scalar of $S$ would however be quite different
because its mass is sensitive to the $U(1)_{\rm PQ}$-breaking terms.
}
within minimal gauge mediation
where the messengers $\Phi+\bar\Phi$ and $\Psi+\bar\Psi$ belong to $5+\bar 5$
representation of the SU$(5)$ into which the SM gauge groups are embedded.
The gauge coupling unification is thus preserved.
To allow mixing between the $\bar 5$ messengers, we simply take the PQ charge
assignment such that $\Psi$ carries a charge opposite to that of $S$ while
all the other messengers are neutral.
The model is then described by the superpotential
\bea
W = W_0(X) + y_\Phi X\Phi\bar\Phi + y_\Psi S\Psi\bar\Psi
+ y_X X\Phi\bar\Psi + y_S S\Psi\bar\Phi,
\eea
in the field basis where the K\"ahler metric is diagonal, ignoring
Planck-suppressed operators.
The above superpotential includes all renormalizable couplings consistent with
the SM gauge invariance and the PQ symmetry.
The effects of hidden sector SUSY breaking are parameterized by a background
singlet field $X$, whose K\"ahler potential should be included to correctly
compute the anomalous dimension of operators depending on $X$.

The PQ symmetry ensures that $S$ corresponds to a flat direction
in the supersymmetric limit for all the messengers fixed at the origin.
Transmitted to the PQ sector by $\Phi+\bar\Phi$, the SUSY breaking effects will
lift this flat direction and fix the vacuum expectation value (VEV) of
the saxion, i.e. the PQ scale.
Meanwhile, the axion remains massless until the QCD instanton effects are turned on.
To examine the saxion potential, it is convenient to construct an effective
theory with messengers integrated out.
Here we assume $|F^X|\ll |X|^2$ so that the mass of $\Phi+\bar\Phi$
is dominated by supersymmetric contribution.
Before proceeding to the analysis, one should note that the $\bar 5$
messengers can be redefined further, without loss of generality, so that either
$X\Phi\bar\Psi$ or $S\Psi\bar\Phi$ is removed in the superpotential
while keeping the K\"ahler metric diagonal.
It is thus natural to assume
\bea
\frac{|y_X|}{|y_\Phi|} + \frac{|y_S|}{|y_\Psi|} = {\cal O}(1),
\eea
in the canonical basis, but $y_\Phi$ and $y_\Psi$ may be hierarchically different
from each other.
The theory has thresholds at scales $\Lambda_\Phi=y_\Phi|X|$ and
$\Lambda_\Psi=y_\Psi|S|$.
For fixed $\Lambda_\Phi$, redefining the $\bar 5$ messengers appropriately,
messenger mixing can be treated perturbatively along the flat direction with
$\Lambda_\Psi$ far from the scale $\Lambda_\Phi$.

\subsection{Saxion potential}

Let us first examine the saxion potential at the region with
$\Lambda_\Psi=y_\Psi|S| \ll \Lambda_\Phi$.
To derive the effective action for $S$, the SUSY breaking messengers can first be
integrated out taking the field basis where the superpotential is written as
\bea
W = W_0(X) + y_\Phi X\Phi\bar\Phi + y_\Psi S\Psi\bar\Psi
+ y_S S \Psi\bar\Phi,
\eea
for the canonical K\"ahler potential.
This approximately corresponds to the mass basis where the term $S\Psi\bar\Phi$
gives small mixing between $\bar 5$ messengers.
The $\Phi+\bar\Phi$, which are integrated out below the scale
$\Lambda_\Phi$, communicate the SUSY breaking to SM-charged fields via gauge
interactions.
In addition, the communication does occur also through the Yukawa interaction
$S\Psi\bar\Phi$ in two ways:
\begin{enumerate}

\item
The anomalous dimensions of $S$ and $\Psi$ are discontinuous at $\Lambda_\Phi$.
As a consequence, the saxion acquires soft mass at the two-loop level
\cite{saxion-run-away-gauge-mediation,Analytic-cont}
\bea
\hspace{-0.8cm}
\frac{m^2_S(\Lambda^-_\Phi)}{M^2_0}
&=& \sum_{\Psi} \left(
\left.\frac{8\pi^2 d y^2_\Psi}{d\ln Q}\right|_{\Lambda^-_\Phi}
-\left.\frac{8\pi^2 d y^2_\Psi}{d\ln Q}\right|_{\Lambda^+_\Phi} \right)
+ \sum_{\Psi,\bar \Phi} \left.\frac{8\pi^2 d y^2_S}{d\ln Q}
\right|_{\Lambda^+_\Phi}
\nonumber \\
&\simeq&
N_\Phi N_\Psi \Big( 5((5N_\Phi N_\Psi+N_\Phi+N_\Psi)y^2_S + y^2_\Phi)
-2(8g^2_3 + 3 g^2_2 + g^2_1 ) \Big)y^2_S,
\eea
where we have neglected the splitting between the Yukawa couplings of
doublet and triplet messengers.
The scalar components of $\Psi+\bar\Psi$ also receive additional soft mass
terms at two loops, besides the ordinary gauge-mediated contributions.

\item
The effective K\"ahler potential for $\Psi$ receives a correction
\bea
\label{K-correction}
\delta K = N_\Phi \frac{y^2_S|S|^2}{y^2_\Phi|X|^2}|\Psi|^2,
\eea
from the tree-level exchange of the SUSY breaking messengers.
Hence, soft mass terms for the scalar components of $\Psi$ receive
additional contribution
\bea
\frac{\delta m^2(\Lambda^-_\Phi)}{M^2_0} =
-N_\Phi(16\pi^2)^2\frac{y^2_S|S|^2}{\Lambda^2_\Phi}.
\eea
The above contribution is always tachyonic and becomes sizable or even
more dominant than the gauge-mediated soft mass for a saxion value
close to $\Lambda_\Phi$.

\end{enumerate}
Here $Q$ denotes the renormalization scale, $N_\Phi$ ($N_\Psi$) is the number
of $\Phi+\bar\Phi$ ($\Psi+\bar\Psi$) pairs, and $M_0$ sets the overall scale
of soft terms generated by the loops of SUSY breaking messengers
\bea
M_0 = -\frac{1}{16\pi^2}\frac{F^X}{X}.
\eea
Note that trilinear couplings for the scalar components of $\Psi+\bar\Psi$
also receive contribution mediated through the Yukawa coupling $y_S$.
The explicit expressions for soft terms are given in the appendix.

With the knowledge of how SUSY breaking is transmitted to the PQ sector,
we further integrate out the remaining messengers $\Psi+\bar\Psi$ under
a large background value of $S$.
The effective action is then determined by the running wave function of $S$
\bea
\label{EFT}
{\cal L}_{\rm eff} = \int d^4\theta\, Z_S(Q=y_\Psi|S|) |S|^2,
\eea
from which the equation of motion for $F^S$ reads
\bea
\label{FS}
\frac{F^S}{S} \simeq -\frac{1}{2}(
\gamma^+_S(\Lambda_\Phi) -\gamma^-_S(\Lambda_\Phi)  )
\frac{F^X}{X}
= -5N_\Phi N_\Psi y^2_S M_0,
\eea
neglecting corrections suppressed by $\Lambda^2_\Psi/\Lambda^2_\Phi$.
Here $\gamma^{\pm}_i$ are the anomalous
dimensions above and below $\Lambda_\Phi$, respectively.
Hence, the scalar potential for the saxion is generated as
\bea
V = V_0 + m^2_S(Q =y_\Psi|S|)|S|^2,
\eea
where $m^2_S$ is the running soft mass of $S$ in the theory between
$\Lambda_\Psi$ and $\Lambda_\Phi$, and a constant $V_0$ has been added
to cancel the cosmological constant.
It should be noted that $m^2_S(Q)$ depends on $|S|$ itself because
the K\"ahler correction (\ref{K-correction}) generates soft terms for
the scalar components of $\Psi+\bar\Psi$ that affect the running of $m^2_S$.
Thus, in the region with $y_\Psi|S|\ll \Lambda_\Phi$, the saxion potential
has a slope approximately given by
\bea
\label{Slop-low-S}
\frac{1}{2M^2_0|S|}\frac{d V}{d|S|} \simeq
\frac{m^2_S(\Lambda^-_\Phi)}{M^2_0}
+ \frac{5N_\Phi N_\Psi}{8\pi^2}\left[  C_\Psi y^2_\Psi
- 2 y^2_S \left(16\pi^2 \frac{y_\Psi|S|}{\Lambda_\Phi} \right)^2 \right]
\ln\left(\frac{y_\Psi|S|}{\Lambda_\Phi}\right),
\eea
where the Yukawa couplings are evaluated at $\Lambda_\Phi$, and
$C_\Psi$ depends on $y^2_{\Psi,S}$ and gauge couplings.
The logarithmic dependence originates from running between $\Lambda_\Phi$
and $\Lambda_\Psi$ through the Yukawa interaction with $\Psi+\bar\Psi$.
Due to the radiative effects from the color-charged scalars of $\Psi+\bar\Psi$,
$C_\Psi$ has a positive value of order unity.

The slope of potential (\ref{Slop-low-S}) shows that the saxion can be
stabilized by the balance between two effects, i.e. the SUSY breaking
mediated by $\Phi+\bar\Phi$ at $\Lambda_\Phi$, and the renormalization effect
through the Yukawa interaction with $\Psi+\bar\Psi$.
The second part of the slope monotonically increases as a function of $|S|$
for $y_\Psi|S|\leq{\cal O}(0.1\Lambda_\Phi)$, and crosses zero at
$y_\Psi|S|={\cal O}(\Lambda_\Phi/8\pi^2)$ because the two contributions
in the bracket have the opposite sign.
Hence, depending on the value of $m^2_S$ at $\Lambda_\Phi$,
the potential develops a minimum along the saxion direction as follows.
If $m^2_S(\Lambda^-_\Phi)$ has a positive value of ${\cal O}(M^2_0)$ or less,
the saxion is stabilized at $y_\Psi|S|\leq {\cal O}(\Lambda_\Phi/8\pi^2)$.
On the other hand, for negative $m^2_S(\Lambda^-_\Phi)$, a minimum
appears at a scale rather close to $\Lambda_\Phi$ where the K\"ahler correction
(\ref{K-correction}) becomes important.
In fixing the VEV of the saxion, the crucial role is played by the messenger
mixing as can be seen from that the potential only has a negative slope
in the limit $y_S\to 0$.

Let's move on to the opposite region with $\Lambda_\Psi\gg \Lambda_\Phi$
along the saxion direction.
In this region, the correct procedure for constructing the effective theory
is to first integrate out the PQ messengers at the scale $\Lambda_\Psi$.
For this, we take the field basis such that
\bea
W = W_0(X) + y_\Phi X\Phi\bar\Phi + y_\Psi S\Psi\bar\Psi
+ y_X X\Phi\bar\Psi,
\eea
for the canonically normalized fields.
Integrating out $\Psi+\bar\Psi$, one obtains the effective action for $X$
determined by its running wave function.
In the effective theory, $S$ does not have any renormalizable interactions,
and the equation of motion for $F^S$ gives
\bea
\frac{F^S}{S} \simeq -5N_\Phi N_\Psi \frac{y^2_X|X|^2}{|S|^2}M_0,
\eea
where corrections suppressed by $\Lambda^2_\Phi/\Lambda^2_\Psi$ have
been neglected.
Hence, the leading contribution to the saxion potential comes from
the dependence on $S$ of the effective wave function of $X$.
Because the anomalous dimension of $X$ is discontinuous at the scale
$\Lambda_\Psi$, the slope of the potential is derived as
\bea
\label{Slop-high-S}
|S|\frac{d V}{d|S|} \simeq \frac{5N_\Phi N_\Psi}{8\pi^2}\left[
y^2_X - \frac{C_{\Phi}}{(8\pi^2)^2} y^2_\Phi \right]|F^X|^2,
\eea
where a positive constant $C_\Phi={\cal O}(g^4_a)$ parameterizes the
contribution induced at the three-loop level.
The potential thus increases at $y_\Psi|S|\gg \Lambda_\Phi$ as a function
of $|S|$, unless $y_X$ is smaller than ${\cal O}(y_\Phi/8\pi^2)$.
This property is cosmologically favorable because the saxion may be
displaced far from the minimum at the end of inflation.
If this happens, the positive slope will make the saxion roll down toward
the true minimum.

The relation (\ref{Slop-high-S}) also gives information about the potential
at saxion values close to $\Lambda_\Phi$, for which messenger mixing can no
longer be treated as a perturbation.
Instead of constructing an effective theory, we use the property that the slope
at $\Lambda_\Psi \gg \Lambda_\Phi$ is positive for $y_X={\cal O}(y_\Phi)$.
This implies that there must exist a minimum below or near $\Lambda_\Phi$
since $m^2_S$ is driven negative at low scales by the Yukawa coupling
with $\Psi+\bar\Psi$.
For the saxion stabilized far below $\Lambda_\Phi$,
the vacuum structure can easily be examined treating the mixing perturbatively.
Note that the potential at $y_\Psi|S|\ll \Lambda_\Phi$ is essentially
determined by $y_{\Psi,S}$ at $\Lambda_\Phi$ and insensitive to the details
of the potential at large $|S|$.
Another possibility is that a minimum lies close to $\Lambda_\Phi$,
which generically requires a rather small $y^2_S$.
The existence of minimum is ensured by the positive slope at
$\Lambda_\Psi\gg \Lambda_\Phi$.

It is worth discussing the situation that there is no mixing between
the messengers, as usually assumed in gauge mediation.
This corresponds to the limit that $y_{X,S}$ vanish.
From the relations (\ref{Slop-low-S}) and (\ref{Slop-high-S}), one then finds
that the potential runs off to infinity along the saxion direction.
Hence, additional SUSY breaking effects are needed to stabilize the saxion.
A natural candidate for this is gravity mediation since the saxion potential
is generated at three-loop level.
Indeed, a higher dimensional operator $\propto |X|^2|S|^2/M^2_{Pl}$ in
the K\"ahler potential gives the gravity-mediated contribution
\bea
\delta V = k\frac{|S|^2}{M^2_{Pl}}|F^X|^2,
\eea
which can compete with the gauge-mediated one to stabilize the saxion
for a positive $k$ of order unity \cite{hadronic-axion-gauge-mediation}.
In the presence of messenger mixing, however, the saxion potential has a slope
as (\ref{Slop-low-S}) at $|S|\ll \Lambda_\Phi$ as long as
$X$ is smaller than ${\cal O}(10^{-3}M_{Pl})$, and the above contribution
becomes important only at $|S|\geq{\cal O}(y_X M_{Pl}/\sqrt{8\pi^2 k})$.

We complete this subsection by summarizing the role of mixing between
SUSY breaking and PQ messengers.
Such mixing indicates that there exist some SM-charged heavy fields that
directly couple both to the SUSY breaking fields and to the PQ breaking field.
It is the SUSY breaking effects transmitted by these fields that
radiatively generate a potential for the saxion and fix the PQ scale.
Moreover, the mixing prevents a runaway behavior of potential at
large saxion values in gauge mediation.
This would be cosmologically relevant for the saxion to settle down
to the true vacuum.

\subsection{Vacuum structure}

Messenger mixing can be treated as a perturbation at $y_\Psi|S| \ll \Lambda_\Phi$
to construct the effective theory for $S$.
Taking into account that $\Psi$ receives a K\"ahler correction (\ref{K-correction})
that contributes to its effective wave function, we examine the vacuum structure
focusing on the case that the saxion is stabilized at
$y_\Psi|S|\leq {\cal O}(\Lambda_\Phi/\sqrt{8\pi^2})$.
From the effective action for $S$, the saxion $\sigma$ and the axino $\tilde a$ are
found to acquire SUSY breaking mass as
\bea
\frac{m^2_\sigma}{M^2_0}
&\simeq&
\frac{5N_\Phi N_\Psi}{4\pi^2} \left[ C_\Psi y^2_\Psi
-4 y^2_S \left(16\pi^2 \frac{y_\Psi S_0}{\Lambda_\Phi}\right)^2
\ln\left(\frac{y_\Psi S_0 }{\Lambda_\Phi}\right) \right],
\nonumber \\
\frac{m_{\tilde a}}{M_0}
&\simeq&
\frac{N_\Psi}{8\pi^2}\left[
3y^2_q \frac{A_q}{M_0} + 2y^2_\ell \frac{A_\ell}{M_0}
+ \frac{5N_\Phi y^2_S}{8\pi^2}
\left(16\pi^2 \frac{y_\Psi S_0}{\Lambda_\Phi}\right)^2
\ln\left(\frac{y_\Psi S_0 }{\Lambda_\Phi}\right) \right],
\eea
for the axion superfield expanded around its VEV,
$S=(S_0+\sigma/\sqrt2)e^{i a/\sqrt2S_0}+\sqrt{2}\theta \tilde a+\theta^2 F^S$.
Here the couplings are evaluated at $Q=y_\Psi S_0$, and
$A_{q,\ell}={\cal O}(10N_\Phi N_\Psi y^2_S M_0)$ are the trilinear couplings
associated with the Yukawa couplings of the PQ triplet and doublet messengers,
respectively.
For $y^2_{\Psi,S}={\cal O}(0.1)$, which are the plausible values, the saxion mass
lies in the range
\bea
{\cal O}\left(\frac{M_0}{\sqrt{8\pi^2}}\right) \leq
m_\sigma
\leq {\cal O}\left(\sqrt{\ln(8\pi^2)} M_0 \right),
\eea
where the upper bound is obtained when
$y_\Psi S_0 ={\cal O}(\Lambda_\Phi/\sqrt{8\pi^2})$.
Though radiatively stabilized, $\sigma$ can be as heavy as the color-charged
MSSM sparticles.
This is because messenger mixing induces a correction to the K\"ahler potential
for $\Psi$ whose loops contribute to the saxion potential.
On the other hand, the axino mass is rather insensitive to the K\"ahler correction
(\ref{K-correction}), and has the value
\bea
m_{\tilde a} = {\cal O}\left(\frac{M_0}{8\pi^2}\right),
\eea
for $y^2_{\Psi,S}={\cal O}(0.1)$, thereby lighter than the MSSM sparticles.

\begin{figure}[t]
\begin{center}
\begin{minipage}{15cm}
\centerline{
{\hspace*{0cm}\epsfig{figure=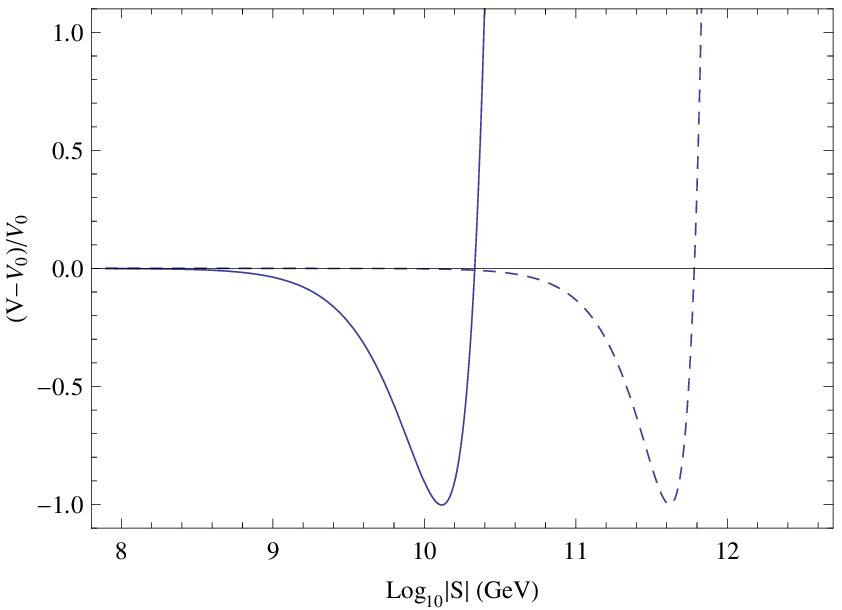,angle=0,width=7.3cm}}
{\hspace*{.2cm}\epsfig{figure=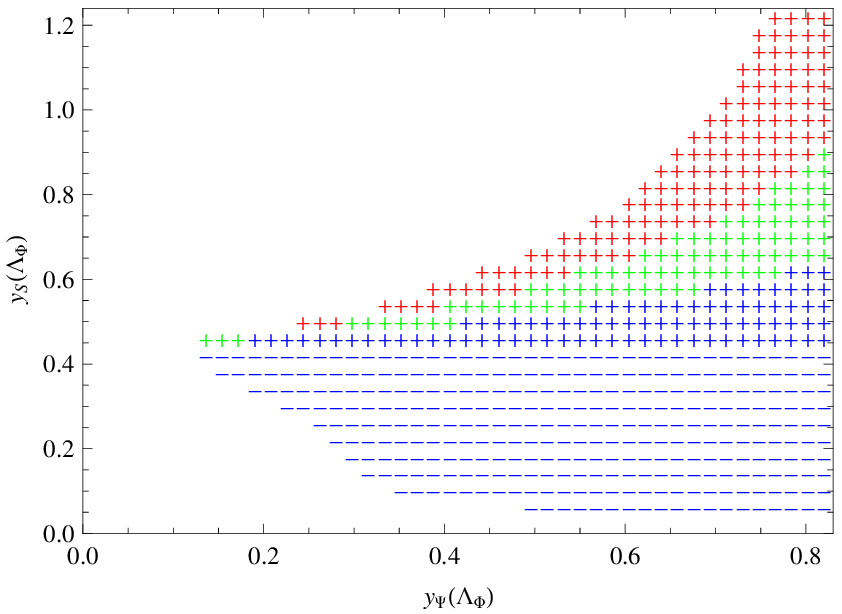,angle=0,width=7.25cm}}
}
\caption{Saxion stabilization in the model with $N_\Psi=1$, $N_\Phi=2$,
$\Lambda_\Phi=10^{13}{\rm GeV}$ and $y_\Phi(\Lambda_\Phi)=0.3$.
The left panel shows the scalar potential along the saxion direction:
the solid line is for the case with $(y_\Psi,y_S) = (0.3,0.5)$ at $\Lambda_\Phi$,
while the dashed one is for $(y_\Psi,y_S) = (0.5,0.4)$.
In the right panel, one can see the dependence of the saxion VEV
on the Yukawa couplings $y_{\Psi,S}$ at $\Lambda_\Phi$:
$S_0=10^{9-10}{\rm GeV}$ (red),
$S_0=10^{10-11}{\rm GeV}$ (green),
$S_0=10^{11-12}{\rm GeV}$ (blue).
The marker $+$ means $m^2_S(\Lambda^-_\Phi)>0$, while $-$
for the case with $m^2_S(\Lambda^-_\Phi)<0$.
}
\label{fig:minimum}
\end{minipage}
\end{center}
\end{figure}

To illustrate how the PQ scale is fixed after SUSY breaking, we provide
some examples.
Fig \ref{fig:minimum} shows the scalar potential along the saxion direction
in the model with $N_\Psi=1$, $N_\Phi=2$, $\Lambda_\Phi=10^{13}$GeV
and $y_\Phi(\Lambda_\Phi)=0.3$.
Depending on the Yukawa couplings $y_{\Psi,S}$ at the scale $\Lambda_\Phi$,
the saxion is stabilized in the following ways:
\bea
\hspace{-0.3cm}
(y_\Psi,y_S) = (0.3,0.5) &:&
S_0\simeq 1.3\times10^{10}{\rm GeV},
\, V_0\simeq (0.4 M_0)^2S^2_0,
\, m_\sigma \simeq 0.5 M_0,
\nonumber \\
\hspace{-0.3cm}
(y_\Psi,y_S) = (0.5,0.4) &:&
S_0\simeq 4.2\times10^{11}{\rm GeV},
\, V_0\simeq (0.9 M_0)^2S^2_0,
\, m_\sigma \simeq 1.7 M_0,
\eea
where $m^2_S(\Lambda^-_\Phi)$ is positive for $(y_\Psi,y_S) = (0.3,0.5)$,
while negative for the other case.
For $m^2_S(\Lambda^-_\Phi)<0$, the saxion potential develops a minimum at a scale near
$\Lambda_\Phi$ while providing a rather large mass to the saxion.
As discussed already, the K\"ahler correction (\ref{K-correction})
to the PQ messenger generates a potential for the saxion that becomes important
at saxion values close to $\Lambda_\Phi$.
In Fig \ref{fig:minimum}, one can also see how the saxion VEV is fixed
depending on $y_{\Psi,S}$ at $\Lambda_\Phi$.
For $y_S<0.42$ where $m^2_S(\Lambda^-_\Phi)$ is negative, the contribution from
the K\"ahler correction (\ref{K-correction}) stabilizes the saxion
at $|S|\geq{\cal O}(10^{-2}\Lambda_\Phi)$.
It is also possible to obtain $S_0={\cal O}(\Lambda_\Phi)$, for instance,
by taking $y_\Psi=0.2$ and $y_S$ smaller than $0.3$.
In this case, the effective theory for $\Psi+\bar\Psi$ constructed by integrating
out $\Phi+\bar\Phi$ is not reliable.
Nonetheless, the relation (\ref{Slop-high-S}) tells that the potential has
a minimum as long as $y_X={\cal O}(y_\Phi)$.

\section{Higgs $\mu$ and $B\mu$ terms}

Although it is an attractive mechanism for generating flavor and CP conserving soft
terms, gauge mediation requires some additional structure to account for
the origin of the $\mu$ and $B\mu$ terms in the MSSM.
In particular, it is quite non-trivial to obtain an acceptable value
of $B$ in theories with gauge mediation.
If one introduces a direct coupling of the Higgs bilinear $H_uH_d$ to the SUSY
breaking field $X$ in the superpotential, one obtains $B={\cal O}(8\pi^2 M_0)$
and thus needs an unnatural fine-tuning to achieve the electroweak symmetry
breaking.
One may instead consider an effective Higgs coupling in the K\"ahler potential
\bea
\int d^4\theta f(X,X^*)H_uH_d,
\eea
which relates $\mu$ to the SUSY breaking parameters\footnote{
This is a generalization of the Giudice-Masiero mechanism
\cite{Giudice-Masiero-mechanism} that generates $\mu$ by SUSY breaking effect.
}.
However, this operator generically gives $B$ of ${\cal O}(8\pi^2 M_0)$ again.
To avoid large $B$, $f$ should have a particular dependence on $X$ such that
generates $\mu$ but not $B\mu$ \cite{small-b-gauge-mediation}, unless there
are other SUSY breaking fields.
For example, one can use $f= X^*/\Lambda$ with some mass scale $\Lambda$
\cite{sweet-spot-susy}.
The dynamics that connects the Higgs sector to the SUSY breaking sector
in such a particular way would generally affect other MSSM soft terms generated
by gauge mediation.

Here, we take an alternative approach to solving both the $\mu$ and $B\mu$
problems, which is provided by the PQ mechanism incorporated
into the gauge mediation.
In fact, a natural solution is to consider a coupling between $H_uH_d$
and the axion superfield $S$.
The $\mu$ term is then induced only after the PQ symmetry is broken.
Furthermore, because the radiative stabilization of the saxion leads to
\bea
\label{FS-B}
\frac{F^S}{S_0} \simeq -5N_\Phi N_\Psi y^2_S M_0,
\eea
a generic coupling of $H_uH_d$ with $S$ is naturally expected to give $B$
of the order of MSSM sparticle masses for $y^2_S={\cal O}(0.1)$\footnote{
A similar idea to suppress $B$ was considered by \cite{B-in-GM} in a
different model.
}.
The above relation is a consequence of mixing between SUSY breaking
and PQ messengers.
In the absence of such mixing, though the saxion can still be stabilized by
adding additional SUSY breaking effects such as gravity mediation,
$F^S/S_0$ would have a value much smaller than $M_0$ since $S$ couples to
$X$ at more than two-loop level.
The relation (\ref{FS-B}) implies that there are simple mechanisms
operative to generate $\mu$ and $B\mu$ terms required for proper
electroweak symmetry breaking:
\begin{itemize}

\item {\it Kim-Nilles} (KN) {\it mechanism}\footnote{
A non-renormalizable superpotential coupling of $H_uH_d$ to the PQ breaking
field has been considered to explain the $\mu$ term in various SUSY breaking
schemes \cite{Kim-Nilles-mechanism}.
The size of $B$ however depends on the mechanism stabilizing the PQ scale.
} :
After the PQ symmetry is broken, the non-renormalizable term
in the superpotential
\bea
{\cal L}_{\rm KN} = \int d^2\theta \lambda\frac{S^2}{M_{Pl}}H_uH_d
+ {\rm h.c.}
\eea
generates $\mu$ and $B\mu$ terms with
\bea
\mu = \lambda \frac{S^2}{M_{Pl}}, \quad
B= -2\frac{F^S}{S}.
\eea
For the saxion stabilized at a scale around $10^{11}$GeV, which is well
within the invisible axion window consistent with astrophysical and cosmological
bounds, $\mu$ is generated at the soft mass scale with $\lambda={\cal O}(0.1)$.
In addition, we obtain $B$ of the correct order of magnitude.

\item {\it Giudice-Masiero} (GM) {\it mechanism} :
For the Higgs fields that couple to $S$ through the effective K\"ahler
potential term
\bea
{\cal L}_{\rm GM} = \int d^4\theta \kappa\frac{S^*}{S}H_uH_d + {\rm h.c.},
\eea
both $\mu$ and $B$ are induced by SUSY breaking effects
\bea
\mu = \kappa \frac{F^{S*}}{S}, \quad B=\frac{F^S}{S}.
\eea
The $\mu$ and $B\mu$ terms are thus of the desired size.
The above coupling in the K\"ahler potential can arise, for instance,
by integrating out heavy fields in the model with the superpotential
terms $\Sigma H_uH_d + \Sigma S S^\prime$ for $S^\prime$ having
a wave-function mixing with $S$ in the K\"ahler potential.

\end{itemize}
It is important to note that, since the phase of $F^S/S_0$ is aligned with $M_0$,
the $B\mu$ term does not introduce new source of CP violation in either mechanism.
Notice also that the Higgs coupling to $S$ fixes the PQ charges of the MSSM
matter fields.
In order for $a$ to play the role of the QCD axion, $U(1)_{\rm PQ}$ should
be anomalous under QCD.
This requires non-zero value for the QCD anomaly coefficient,
$N=N_\Psi \pm 6$, $+$ for the GM while $-$ for the KN mechanism\footnote{
Models that incorporate the KN mechanism with $N=\pm1$ are free from
the domain wall problem.
For example, in the model with $N_\Psi=5$ and $N_\Phi=1$,
one obtains $N=-1$, and the MSSM gauge couplings remain perturbative up
to the unification scale as long as
$10^6{\rm GeV}\lesssim \Lambda_\Psi\leq 0.1 \Lambda_\Phi$.
Meanwhile, for the case that $\mu$ is generated by the KN or GM mechanism
with $N\neq \pm 1$, one can consider the situation that the saxion is
displaced far from the origin during the inflation.
Then, the reheating would not restore $U(1)_{\rm PQ}$ after inflation.
In addition, provided that the fluctuation around the initial displacement
due to the quantum fluctuation during the de-Sitter expansion is small
enough, the saxion will settle down to one of the $|N|$ degenerate vacua.
This will provide a solution to the domain wall problem
(see \cite{domain-wall}, for a similar consideration).
}.

Another important consequence of the Higgs coupling to $S$ is that mixing
between the saxion (axino) and neutral Higgs (Higgsino) fields is induced
after electroweak symmetry breaking.
Through the mixing, the saxion and axino interact also with other MSSM particles.
The saxion mixing term with neutral Higgs bosons is obtained from
\bea
\label{mixing-1}
{\cal L}_{\rm Higgs} = -|\mu|^2 (|H^0_u|^2+|H^0_d|^2 )
+ |B \mu| (H^0_u H^0_d + {\rm c.c.}),
\eea
by making the replacement
\bea
|\mu| \rightarrow \frac{{\cal C}_\sigma|\mu|}{S_0}
\frac{\sigma}{\sqrt2} \quad{\rm with}\quad
{\cal C}_\sigma = \left.\frac{\partial\ln |\mu|}{\partial\ln|S|}
\right|_{S=S_0},
\eea
where we have used that $F^S/S$ does not depend on $S$, as can be seen
from (\ref{FS}).
Therefore, the saxion slightly mixes with the neutral CP even Higgs bosons
through the interaction suppressed by $v/S_0$ with
$v^2=\langle |H^0_u| \rangle^2 + \langle |H^0_d| \rangle^2$.
Similarly, the axino has tiny mixing with the neutral Higgsinos
determined by
\bea
\label{mixing-2}
{\cal L}_{\tilde a} =
\frac{{\cal C}_{\tilde a}\mu}{S_0}
( H^0_u \tilde H^0_d + H^0_d \tilde H^0_u )\tilde a
+ {\rm h.c.},
\eea
with
\bea
{\cal C}_{\tilde a} =
\left.\frac{\partial \ln \mu}{\partial \ln S}\right|_{S=S_0}.
\eea
Note that the coefficient ${\cal C}_\sigma$ crucially depends on the mechanism
for generating the $\mu$ term:
\bea
{\cal C}_\sigma|_{\rm KN} =2,
\quad
{\cal C}_\sigma|_{\rm GM} = Q\frac{d \ln \kappa}{d Q}
\Big|_{Q=y_\Psi S_0}
= {\cal O}\left(\frac{1}{8\pi^2}\right),
\eea
whereas the size of ${\cal C}_{\tilde a}$ is insensitive to the form
of the Higgs coupling to $S$:
\bea
{\cal C}_{\tilde a}|_{\rm KN}=2, \quad
{\cal C}_{\tilde a}|_{\rm GM}=-1.
\eea

Let us examine the saxion/axino couplings to the MSSM sector.
First, there are the interactions,
$\sigma|H^0_u||H^0_{u,d}| + \sigma\tilde H^0_u \tilde H^0_d +
H^0_{u,d}\tilde H^0_{d,u}\tilde a$, that are induced directly from
the Higgs coupling to $S$.
The couplings for these interactions are non-vanishing even in the limit $v\to 0$.
To derive other interactions between the saxion/axino and the MSSM particles,
the small mixing terms from (\ref{mixing-1}) and (\ref{mixing-2}) should be
removed by performing an appropriate field redefinition.
Indeed, the saxion and axino couplings can be read off from the MSSM
Lagrangian by the substitutions
\bea
\label{replacement-saxion}
H^0_{d,u} &\to&
-\frac{{\cal C}_\sigma v}{S_0}
\frac{|\mu|^2}{m^2_h-m^2_\sigma} \frac{N^\sigma_{d,u} \sigma}{\sqrt2}  ,
\\
\label{replacement-axino}
(\tilde B,\tilde W^0,\tilde H^0_d,\tilde H^0_u) &\to&
-\frac{{\cal C}_{\tilde a} v}{S_0}
N^{\tilde a}_{\tilde B,\tilde W^0,\tilde H^0_d,\tilde H^0_u} \tilde a,
\eea
where $m_h$ is the mass of the lightest CP even neutral Higgs boson $h$, and
$N^{\tilde a}_{\tilde B,\tilde W^0}$ are non-vanishing because $\tilde H^0_{u,d}$
mix with the bino and neutral wino.
The mixing parameters are presented in the appendix.
After the replacement, one must diagonalize the mass matrix for
$(\tilde B,\tilde W^0,\tilde H^0_d,H^0_u)$.
In addition to those proportional to ${\cal C}_{\sigma,\tilde a}$,
interactions between the saxion/axino and MSSM particles are also generated
by the loops involving PQ messengers that acquire mass from the VEV of $S$.
These couplings can be derived from the dependence on $S$ of the MSSM gauge
couplings
\bea
\label{S-dependence}
\frac{1}{g^2_a} = -\frac{N_\Psi}{16\pi^2}\ln(S^*S) + (S\mbox{-independent part}),
\eea
in the effective theory with $\Psi+\bar\Psi$ integrated out.
The saxion decay to MSSM particles and the decay of heavy sparticles into axino
will be discussed later, after examining MSSM sparticle masses.

We stress that the axion superfield can play an important role not only in solving
the strong CP problem but also in explaining the presence of the $\mu$ and $B\mu$
terms within theories with gauge mediation.
This nice feature stems from the mixing between SUSY breaking and PQ messengers,
through which the SUSY breaking is communicated to the PQ sector and radiatively
stabilizes the PQ scale.
In fact, an appropriate PQ charge assignment is the only thing that was needed
to allow such mixing between the messengers.

\section{Phenomenological implications}

In this section, we discuss the phenomenological aspects of the model.

\subsection{Sparticle masses}

To derive soft terms for the MSSM fields, we begin by summarizing the possible
range of threshold scales in the theory.
The PQ scale, which is radiatively stabilized in the presence of mixing between
messengers, is constrained by various astrophysical and cosmological
observations \cite{review-axion}.
On the other hand, an upper bound is put on $\Lambda_\Phi=y_\Phi|X|$ to suppress
gravity mediation that in general generates flavor-violating soft terms with
size $F^X/M_{Pl}$.
These constraints lead to
\bea
\label{Scales}
10^9{\rm GeV}\lesssim S_0 \lesssim 10^{12}{\rm GeV}, \quad
10S_0 \lesssim |X| < 10^{15}{\rm GeV},
\eea
where the lower bound on $X$ has been put to concentrate on the case
that messenger mixing can be treated perturbatively.
Since the theory contains heavy messengers, the MSSM soft terms at TeV scale are
determined by the parameters
$\{M_0,\Lambda_\Phi,N_\Phi,\Lambda_\Psi,N_\Psi\}$, while $\mu$ and $B$
are generated by the KN or GM mechanism.

At the higher threshold $\Lambda_\Phi$, the SUSY breaking is transmitted to
the MSSM sector by $\Phi+\bar\Phi$ through ordinary gauge
mediation \cite{gauge-mediation}.
The threshold effects induced by these messengers generate gaugino and scalar
soft masses as
\bea
\frac{M_a(\Lambda^-_\Phi)}{M_0} &=& N_\Phi g^2_a(\Lambda_\Phi),
\nonumber \\
\frac{m^2_i(\Lambda^-_\Phi)}{M^2_0} &=&
2N_\Phi C^a_i g^4_a(\Lambda_\Phi),
\eea
for $C^a_i$ being the quadratic Casimir of the corresponding field.
Soft trilinear terms arise at two-loop order, and thus are negligible
at $\Lambda_\Phi$.
Since the radiative corrections due to the SM gauge interaction
play the dominant role for the mediation, soft terms preserve flavor and CP.
The gauge-mediated soft parameters are subsequently RG evolved down to
$\Lambda_\Psi=y_\Psi S_0$ in the presence of SM-charged fields $\Psi+\bar\Psi$.

Low energy soft terms below $\Lambda_\Psi$ can be computed integrating
out $\Psi+\bar\Psi$.
Since the saxion is stabilized with $F^S/S_0={\cal O}(M_0)$, the PQ
messengers induce negligible threshold effects for gaugino masses and trilinear
couplings.
However, depending on the saxion VEV, soft mass terms for the MSSM scalar
fields can receive sizable threshold corrections.
This is because the PQ scalars acquire soft mass also from the K\"ahler correction
(\ref{K-correction}).
The threshold effect provides flavor-universal soft scalar masses
\bea
\frac{\Delta m^2_i(\Lambda^-_\Psi)}{M^2_0} \simeq
4N_\Phi N_\Psi\left[ C^a_i g^4_a
-Y_i g^2_Y R \right] \frac{y^2_S S^2_0}{\Lambda^2_\Phi}
\ln\left(\frac{\Lambda_\Phi}{\Lambda_\Psi}\right),
\eea
where the gauge couplings are evaluated at $Q=\Lambda_\Psi$, and
corrections suppressed by the two-loop factor have been neglected.
The above contribution includes the hypercharge trace term, which
is the part proportional to the hypercharge $Y_i$.
This term is non-vanishing due to the Yukawa splitting of doublet and
triplet components in $\Psi+\bar\Psi$:
\bea
R = 8\pi^2\frac{y^2_{S_q}(\Lambda_\Phi)-
y^2_{S_\ell}(\Lambda_\Phi)}{y^2_S(\Lambda_\Phi)} \sim
\ln\left(\frac{M_{\rm GUT}}{\Lambda_\Phi}\right),
\eea
with $y^2_S=(y^2_{S_q}+ y^2_{S_\ell})/2$, and $M_{\rm GUT}$ being the
unification scale.
The other term in the bracket arises from the loops of $\Psi+\bar\Psi$ since
the supertrace of their mass matrix is non-vanishing
\cite{non-standard-gauge-mediation}.
For example, depending on the values of $y_{\Psi,S}$ at $M_{\rm GUT}$, the model
with $N_\Phi=N_\Psi=1$ and $\Lambda_\Phi=10^{12}$GeV leads to
\bea
(y_\Psi,y_S)=(0.34,0.68) &:& R\simeq 14.5,\, S_0 \simeq 3.5\times10^{9}{\rm GeV},
\nonumber \\
(y_\Psi,y_S)=(0.24,0.48) &:& R\simeq 14.3,\, S_0 \simeq 8.7\times10^{10}{\rm GeV},
\eea
where we have assumed $y_\Phi\ll 1$ to evaluate the running of $y_{\Psi,S}$
from $M_{\rm GUT}$ to $\Lambda_\Phi$.
In the case with $(y_\Psi,y_S)=(0.24,0.48)$ at $M_{\rm GUT}$,
the PQ scale is close to $\Lambda_\Phi$, and thus the MSSM scalars receive
non-negligible threshold correction $\Delta m^2_i \simeq -0.13 Y_i M^2_0$
at $\Lambda_\Psi\simeq 2\times 10^{10}$GeV.
It is worth noting that one can simply change the sign of $R$ by considering
mixing between $5$ messengers instead of between $\bar 5$ ones.
Hence, $R$ can be of either sign.

The threshold effect at $\Lambda_\Psi$ for soft scalar masses becomes important
when the saxion VEV is close to $\Lambda_\Phi$, as can be deduced from its origin.
For instance, if $y_\Psi S_0={\cal O}(\Lambda_\Phi/\sqrt{8\pi^2})$, the hypercharge
trace term makes $\Delta m^2_i$ comparable to the gauge-mediated soft mass
for scalars that are charged only under $U(1)_Y$.
This indicates that the lightest ordinary sparticle (LOSP) can be provided
by the stau even for small $N_\Phi$, if $R$ is positive and the saxion is stabilized
at a scale near $\Lambda_\Phi$.
However, for the saxion VEV much lower than $\Lambda_\Phi$, the PQ messengers
only give negligible threshold to soft terms.
Meanwhile, the gauge-mediated sfermion masses satisfy two sum-rules
\bea
\sum_i Y_i m^2_i = \sum_i (B_i-L_i) m^2_i =0,
\eea
where the sum is over one generation of sfermions, and $B$ ($L$) is the
baryon (lepton) number.
Affected by the PQ threshold $\Delta m^2_i$, these sum rules can be used to extract
information about the PQ sector.

Finally, we examine which particle is the LSP.
The theory contains two light fermionic sparticles, the axino and the gravitino.
To determine the gravitino mass, we assume that $X$ is the goldstino superfield
whose $F$-term cancels the cosmological constant.
The gravitino then absorbs the fermionic component of $X$, $\tilde X$, to become
massive with
\bea
\label{gravitino-mass}
\frac{m_{3/2}}{M_0} = \frac{16\pi^2}{\sqrt3}\frac{X}{M_{Pl}},
\eea
which lies in the range $10^{-6}M_0\lesssim m_{3/2} < 10^{-1}M_0$ for the theory
with (\ref{Scales}).
Using that the axino acquires mass of ${\cal O}(M_0/8\pi^2)$ for
$y^2_{\Psi,S}={\cal O}(0.1)$, one also arrives at the relation
\bea
\label{LSP}
\frac{m_{3/2}}{m_{\tilde a}} \sim \frac{X}{10^{14}{\rm GeV}}.
\eea
This shows that the axino becomes the LSP if $X\geq {\cal O}(10^{14})$GeV.
The LSP would otherwise be given by the gravitino.
Though we will not consider it here, there is a possibility to have an axino
LSP even for $X<10^{14}$GeV.
One way is to consider $y^2_{\Psi,S}\ll 0.1$.
For instance, in models with $y^2_{\Psi,S}={\cal O}(10^{-2})$, we obtain
$m_{\tilde a}={\cal O}(10^{-3}M_0)$ and $m_\sigma\geq{\cal O}(0.1M_0)$.
For $y^2_S\ll 0.1$, a Higgs coupling to $S$ will give $|B| \ll {\cal O}(M_0)$,
with which it is still possible to achieve the correct electroweak symmetry
breaking.
One can also consider other models where $\tilde X$ is not the main component
of the goldstino.
Then, $F^X$ will have a VEV less than ${\cal O}(m_{3/2}M_{Pl})$, and thus
the gravitino can be heavier than the axino for $X<10^{14}$GeV.
In this case, one would need $m_{3/2}\ll m_\sigma$ in order not to destabilize
the PQ scale, because gravity mediation provides soft mass typically
of ${\cal O}(m_{3/2})$ to the saxion as well as to the MSSM scalars.

\subsection{Decay of sparticles}

The heavy MSSM sparticles rapidly decay into the LOSP, denoted by $\tilde \chi$,
which subsequently decays into
lighter sparticles, i.e. into the axino or gravitino.
The decay of $\tilde\chi$ occurs more slowly because the axino and gravitino
are very weakly coupled to other particles.
Here we are assuming R-parity conservation.
Measurement of the decay length of $\tilde\chi$ will give direct information either
about the SUSY breaking scale or about the PQ scale, depending on which of axino
and gravitino is the main decay product\footnote{
However, the decay length alone would not allow us to distinguish between
SUSY breaking scenarios where the LSP is given either by the axino or
by the gravitino \cite{axino-gravitino-collider}.
}.
In fact, the axino and gravitino have similar type of interactions that mediate
the decay of heavy sparticles.

At energy scales much higher than $m_{3/2}$, the gravitino $\tilde G$
effectively behaves as a goldstino.
To study the decay of heavy sparticles into gravitino,
the effective interaction Lagrangian for the goldstino component
can be written in non-derivative form \cite{effective-gravitino-interaction}.
Using the relation $F^X=-16\pi^2M_0X\simeq \sqrt{3}m_{3/2}M_{Pl}$,
one obtains
\bea
{\cal L}^{\tilde G}_{\rm int} =
\frac{i}{16\pi^2 X}\left(
\frac{m^2_\phi-m^2_\psi}{M_0} \phi^* \psi \tilde X
+ \frac{1}{4\sqrt 2}\frac{M_\lambda}{M_0}
\tilde X \sigma^{\mu\nu} \lambda F_{\mu\nu}\right) + {\rm h.c.},
\eea
independently of how the SUSY breaking is mediated.
Here $\lambda$ stands for the gaugino, and $F_{\mu\nu}$ is the corresponding
field strength, while $\phi$ is the scalar, and $\psi$ is its fermionic partner.
The gravitino interactions are proportional to the mass splitting in
the supermultiplet, and inversely proportional to $F^X$.

Since the $\mu$ term is generated after the PQ symmetry breaking,
${\cal C}_{\tilde a}$ has a value of order unity, and the axino interacts with MSSM
particles through
\bea
H^0_{u,d}\tilde H^0_{d,u}\tilde a +
\left(H^{0*}_{u,d}\tilde B \tilde H^0_{u,d}
+ \tilde f^c f \tilde H^0_{u,d}+ \tilde f^* \tilde B f +
\tilde f^* \tilde W^0 f
+ \tilde H^0_{u,d}\sigma^\mu \bar{\tilde H}^0_{u,d} Z_\mu\right),
\eea
where we have omitted the couplings.
For those in parenthesis, the axino couplings are obtained after the replacement
(\ref{replacement-axino}), and thus would vanish in the limit $v\to 0$.
Here $Z_\mu$ is the Z-boson with mass $M_Z$, while $f$ denotes the SM fermion,
and $\tilde f$ is its scalar partner.
In addition, there are axino interactions induced at the loop level
\bea
H^{0*}_{u,d}\tilde H^0_{u,d} \tilde a
+ \tilde f^* f \tilde a + \tilde a \sigma^{\mu\nu}\lambda F_{\mu\nu}.
\eea
whose couplings are determined by the dependence on $S$ of the gauge couplings
after integrating out $\Psi+\bar\Psi$.
Because of the axino mixing with $\tilde H^0_{u,d}$, the above couplings also
receive contribution from the loops involving MSSM particles that become massive
after the electroweak symmetry breaking.

Using the Lagrangian for axino/gravitino interactions, one can estimate
the lifetime of $\tilde\chi$, which is also subject to cosmological constraints.
In the model, the LOSP can be provided by the stau or bino.
Let us first examine the case that the bino is the LOSP.
The coupling for the interaction $h\tilde B\tilde a$ receives contribution
from that of $H^{0*}_{u,d}\tilde B \tilde H^0_{u,d}$ due to the axino component of
neutral Higgsinos, and from $H^0_{u,d}\tilde H^0_{d,u}\tilde a$ through
the mixing between $\tilde B$ and $\tilde H^0_{u,d}$.
This coupling is of ${\cal O}(M_Z/S_0)$, and thus the bino mainly
decays into $h$ and $\tilde a$ with decay width
\bea
\Gamma_{\tilde B\to h\tilde a} \sim
\frac{1}{10^{-8}{\rm sec}}
\left(\frac{M_{\tilde B} }{200{\rm GeV}}\right)
\left(\frac{10^{10}{\rm GeV}}{S_0}\right)^{2},
\eea
for $M_{\tilde B}>m_h+m_{\tilde a}=m_h+{\cal O}(M_0/8\pi^2)$, assuming that
other neutral Higgs bosons are very heavy.
It is easy to see that the bino decay into gravitino is highly suppressed.
For a bino with $M_{\tilde B}<m_h+m_{\tilde a}$, the dominant decay channel is
$\tilde B\to Z\tilde a$.
The coupling for this decay process is additionally suppressed by ${\cal O}(M_Z/\mu)$
compared to that of $h\tilde B\tilde a$, because it requires
bino-Higgsino mixing as well as the axino-Higgsino mixing.
As it is mediated by the interaction induced at the loop level, the decay
$\tilde B\to\tilde a \gamma$ has a small branching ratio.
On the other hand, in the case of a stau LOSP, the decay takes place
via the interaction $\tilde \tau^* \tau \tilde a$ with coupling
of ${\cal O}((M_Z N^{\tilde a}_{\tilde B}
+ m_f N^{\tilde a}_{\tilde H^0_d}/\cos\beta)/S_0)$.
The decay rate is estimated as
\bea
\hspace{-0.5cm}
\Gamma_{\tilde \tau \to \tau \tilde a} \sim
\frac{1}{10^{-7}{\rm sec}}\left[
\left(\frac{M_Z\cos2\beta}{0.3 M_{\tilde B}}\right)^2
+ \frac{(1-n_{\tilde a}\tan\beta)^2}{10^2} \right]
\left(\frac{m_{\tilde \tau}}{200{\rm GeV}}\right)
\left(\frac{10^{10}{\rm GeV}}{S_0}\right)^{2},
\eea
with $n_{\tilde a}={\cal O}(M^2_Z/\mu M_{\tilde B})$, and
$\tan\beta=\langle |H^0_u| \rangle/\langle |H^0_d| \rangle$.
Therefore, for a bino or stau LOSP, the LOSP will decay mainly into axinos,
rather than into gravitinos.
Measuring its decay length, one can thus extract information about the PQ scale.
It is interesting to note that, depending on the PQ scale, a bino LOSP can
decay inside the detector while leaving displaced vertices.
For a stau LOSP, there is a possibility that decaying staus can appear in the
detector often enough to measure their charged tracks.
Since $\tilde\chi$ has a lifetime much shorter than a second, the nucleosynthesis
does not place any significant bound.

The LSP, into which all the sparticles eventually decay, can be either the axino
or the gravitino depending on the scale of gauge mediation, and constitutes
the dark matter of the Universe.
If it is lighter than the gravitino, the axino becomes a good candidate for
the cold dark matter \cite{axino-dark-matter,axino-cold-dark-matter}.
The axino can become the LSP when $X\geq{\cal O}(10^{14})$GeV for
$y^2_{\Psi,S}={\cal O}(0.1)$.
In this case, the gravitino decay occurs through the interaction
\bea
{\cal L}_{\rm int} = \frac{i}{2 M_{Pl}}\bar{\tilde a} \gamma^\mu\gamma^\nu
\tilde G_\mu\partial_\nu a + {\rm h.c.},
\eea
which leads to
\bea
\Gamma_{\tilde G\to \tilde a a}
\sim
\frac{1}{0.5\times 10^{13}{\rm sec}}
\left(\frac{8\pi^2m_{3/2}}{M_0}\right)^3
\left(\frac{M_0}{500{\rm GeV}}\right)^3.
\eea
The gravitino decay will thus produce an axino and axion.
On the other hand, if heavier than the gravitino, axinos produced by the decay
of $\tilde\chi$ will decay into gravitino.
Since it is mediated by the effective interaction
\bea
{\cal L}^{\rm eff}_{\rm int} =
\frac{i}{8\pi^2 X}\frac{m_{\tilde a}}{M_0}
\tilde X \sigma^\mu \bar{\tilde a} \partial_\mu a + {\rm h.c.},
\eea
for $m_{\tilde a}>m_{3/2}$, the decay $\tilde a\to \tilde G a$ occurs with
\bea
\Gamma_{\tilde a \to \tilde G a} \sim
\frac{1}{0.5\times 10^{13}{\rm sec}}
\left(\frac{m_{\tilde a}}{m_{3/2}}\right)^2
\left(\frac{8\pi^2m_{\tilde a}}{M_0}\right)^3
\left(\frac{M_0}{500{\rm GeV}}\right)^3,
\eea
where we have used the relation (\ref{gravitino-mass}).
The axino will decay with a long lifetime, producing gravitinos together
with axions.
Note that late decay of axino/gravitino produces LSPs and axions, which
may be warm or even hot at present unless $\tilde a$ and $\tilde G$ are
highly degenerate in mass.
In fact, having a free-streaming length much larger than
${\cal O}(10){\rm Mpc}$, LSPs produced by such late decays will behave like
a hot dark matter.
The energy density of hot dark matter is severely constrained by the CMBR
and structure formation \cite{CMB,hot-dark-matter}.
We will return to this issue in the next section.

\section{Cosmological aspects}

The theory contains the saxion that has a rather flat potential generated
after SUSY breaking and interacts with other particles with coupling
suppressed by the PQ scale.
This scalar may play some non-trivial role in cosmology as its potential can
receive additional sizable contribution at early Universe.
The relic abundance of dark matter depends on the cosmological evolution
of the saxion.
It is thus of importance to understand the saxion properties.

\subsection{Saxion decay}

Because it acquires mass as ${\cal O}(0.1M_0)\leq m_{\sigma}\leq {\cal O}(M_0)$
depending on $S_0/\Lambda_\Phi$, the saxion will decay into axino, gravitino,
and light MSSM particles.
The interaction relevant for its decay can be derived from the effective
action (\ref{EFT})
\bea
\label{saxion-lsp}
{\cal L}^\sigma_{\rm int} \simeq
\frac{\sigma}{\sqrt2 S_0}\left[
(\partial^\mu a)\partial_\mu a
+ \left(
\frac{c_{\tilde a}}{2} m_{\tilde a}\tilde a\tilde a
+ \frac{1}{16\pi^2}\frac{m^2_\sigma}{M_0}\frac{S_0}{X} \tilde a\tilde X
+ {\rm h.c.} \right)
\right],
\eea
where we have included the effective interaction with the goldstino.
The coefficient $c_{\tilde a}=(\partial\ln m_{\tilde a}/\partial \ln|S|)|_{S=S_0}$
has a value of ${\cal O}(0.1)$ or less for
$y^2_\Psi S^2_0\leq {\cal O}(10^{-3}\Lambda_\Phi)$.
If the saxion is stabilized at $y_\Psi|S|={\cal O}(0.1\Lambda_\Phi)$,
the K\"ahler correction (\ref{K-correction}) becomes important and leads
to $c_{\tilde a}={\cal O}(1)$ and $m_\sigma={\cal O}(M_0)$.
Besides the above interactions, the saxion also has couplings to MSSM particles
when the $\mu$ term arises after PQ symmetry breaking:
\bea
\label{saxion-int1}
\sigma \tilde H^0_u \tilde H^0_d + \sigma |H^0_u| |H^0_{u,d}| +
\left( H^0_{u,d}f\bar f + H^{0*}_{u,d}\tilde B \tilde H^0_{u,d}
+ H^0_{u,d}\tilde f \tilde f^c + H^0_{u,d} Z^\mu Z_\mu + \cdots
\right),
\eea
where the first two terms come from the Higgs coupling to $S$, while the other
interactions require the saxion component of $H^0_{u,d}$.
The ellipsis contains the interactions with $W^{\pm}_\mu$ and $\tilde W^{0,\pm}$.
There are also saxion interactions induced at the loop level,
which include
\bea
\label{saxion-int2}
\sigma \tilde f \tilde f^* + \sigma F^{\mu\nu} F_{\mu\nu},
\eea
where the couplings can be derived using the relation (\ref{S-dependence}),
and making the replacement (\ref{replacement-saxion}).
Though it is suppressed by the loop factor, the interaction
$\sigma \tilde f \tilde f^*$ can be important because
the saxion coupling from $H^0_{u,d}\tilde f\tilde f^c$ is suppressed
by ${\cal C}_\sigma A_f/M_0={\cal O}({\cal C}_\sigma/8\pi^2)$
in gauge mediation.
Here $A_f$ is the soft trilinear coupling.

Before examining the partial decay widths of the saxion, we note that the
color-charged
sparticles are much heavier than uncolored sparticles in minimal gauge mediation.
This results in that the electroweak symmetry breaking is achieved under a fine
tuning of a few percent between the Higgs mass parameters.
So we shall consider the $\mu$ term with\footnote{
Using that the combination $M_a/g^2_a$ is invariant under the RG
evolution at the one-loop level, one obtains $M_{\tilde B}\simeq 0.22N_\Phi M_0$
at $Q=1$TeV.
The bino mass is thus about $100$GeV for $N_\Phi=1$ and $M_0=500$GeV.
In the subsequent discussion, we will treat $M_{\tilde B}$ as a free
parameter satisfying $M^2_{\tilde B}\ll M^2_0$.
}
\bea
\frac{m^2_h}{M^2_0}, \frac{m^2_{\tilde \tau}}{M^2_0},
\frac{M^2_{\tilde B}}{M^2_0} \ll \frac{|\mu|^2}{M^2_0} \leq {\cal O}(1).
\eea
For later discussion, it is convenient to define the total decay width
of the saxion as
\bea
\Gamma_\sigma \equiv \frac{1}{64\pi B_a}\frac{m^3_\sigma}{S^2_0},
\eea
with $B_a$ being the branching ratio into axions.
Let us now examine the saxion decay.
From (\ref{saxion-lsp}), the decay widths for $\sigma\to \tilde a \tilde a$
and $\sigma\to a\tilde G$ are determined by
\bea
\label{saxion-to-LSPs}
(\Gamma_{\sigma\to \tilde a\tilde a},\Gamma_{\sigma\to \tilde a\tilde G})
\simeq
B_a\Gamma_\sigma
\left(\frac{c^2_{\tilde a}}{2} \frac{m^2_{\tilde a}}{m^2_\sigma},
\frac{1}{(8\pi^2)^2}\frac{m^2_\sigma}{M^2_0}\frac{S^2_0}{X^2}
\right).
\eea
The saxion also decays into MSSM particles via the interactions
(\ref{saxion-int1}) and (\ref{saxion-int2}), and thus crucially depends on the
form of the Higgs coupling to $S$.
In particular, for ${\cal C}_\sigma={\cal O}(1)$, the saxion decay occurs
with
\bea
\label{saxion-decay-width}
(\Gamma_{\sigma\to hh},\Gamma_{\sigma\to f \bar f},\Gamma_{\sigma\to ZZ})
\simeq
4{\cal C}^2_\sigma B_a \Gamma_\sigma \frac{|\mu|^4}{m^4_\sigma}\left(
k^2_h,
\frac{N_f k^2_f m^2_f m^2_\sigma}{(m^2_h-m^2_\sigma)^2},
\frac{k^2_Z M^4_Z}{(m^2_h-m^2_\sigma)^2} \right),
\eea
if the processes are kinematically allowed, where
$k_h=1+|B|\sin2\alpha/2|\mu|$, $k_Z=N^\sigma_u \sin\beta+N^\sigma_d \cos\beta$,
and we have neglected the mass of the decay products.
In the decay width into the SM fermions, $N_f=3\,(1)$ for quarks (leptons), and
$k_f=N^\sigma_u/\sin\beta$ for up-type quarks while $k_f=N^\sigma_d/\cos\beta$
for down-type quarks and leptons.
Here $k_{h,f,Z}$ have a value of order unity.
Furthermore, if it is heavy enough, the saxion will decay into MSSM sparticles
with the decay widths roughly estimated as
\bea
(\Gamma_{\sigma\to \tilde B\tilde B},
\Gamma_{\sigma\to \tilde \tau \tilde \tau^*})
\sim
4 B_a\Gamma_\sigma
\frac{|\mu|^2}{m^2_\sigma}
\left(
\frac{{\cal C}^2_\sigma M^4_Z}{(m^2_h-m^2_\sigma)^2},
\frac{N_\Psi k^2_{\tilde \tau}}{(8\pi^2)^2} \frac{M^4_0}{m^2_\sigma|\mu|^2}
\right),
\eea
where $\tilde B$ is a bino-dominant neutralino with small Higgsino component,
and the stau has
$k_{\tilde \tau} =4N^2_\Phi C^a_{\tilde \tau}(g^6_a(\Lambda_\Psi)-g^6_a(M_0))/b_a$.
To estimate the decay width into MSSM sparticles, we have used that the interaction
$\sigma\tilde B\tilde B$ arises from $\sigma \tilde H^0_u \tilde H^0_d$
and $H^{0*}_{u,d}\tilde B\tilde H^0_{u,d}$ due to the bino-Higgsino mixing,
and that gauge mediation gives $A_\tau={\cal O}(M_0/8\pi^2)$.

\subsection{Cosmological constraints}

During the inflationary epoch, the saxion is displaced from the true
vacuum because it obtains a Hubble-induced mass term.
After inflation, as the Universe is reheated, the PQ messengers generate
a thermal potential for the saxion, $\delta V \sim y^2_\Psi T^2|S|^2$ at
$|S|\ll T$.
For $|S|\gg T$, the messengers become massive and only give small thermal
effect.
Hence, if it sits around the origin just after inflation, the saxion is
thermally trapped at the origin until the temperature drops down
to $T={\cal O}(M_0)$.
This implies that thermal inflation \cite{thermal-inflation} occurs when
the potential energy $V_0\sim m^2_\sigma S^2_0$ dominates the Universe.
After thermal inflation, the saxion begins to oscillate about the true
minimum with an amplitude of ${\cal O}(S_0)$.
It is also possible that the saxion is shifted far from the origin during
primordial inflation.
In this case, the coherent oscillation of the saxion starts with an amplitude
less than $M_{Pl}$ when the Hubble parameter becomes comparable to $m_\sigma$.
Since it behaves like matter, the oscillation energy would dominate the energy
of the Universe if the initial amplitude is large enough.

Let us examine cosmological constraints of the saxion properties
in the case that the saxion dominates the Universe at an early time,
which is a plausible possibility as discussed above.
Since a late-time entropy production would alleviate the constraints placed
on the saxion, we further assume that there is no additional entropy generation
after the saxion decay.
An important constraint then comes from the axions produced by the saxion decay
as they behave like neutrinos \cite{axions-from-saxion-decay}.
If there is no coupling between $H_uH_d$ and $S$, the saxion mainly
decays into two axions with $B_a\simeq 1$.
Obviously, the produced axions would spoil the Big Bang nucleosynthesis.
However, the situation can be different if the $\mu$ term is generated
after $U(1)_{\rm PQ}$ breaking.
From (\ref{saxion-decay-width}), one sees that the saxion couplings
to the MSSM particles can easily suppress $B_a$ below $0.1$ for
\bea
m^2_\sigma \ll |\mu|^2,
\eea
when the $\mu$ term is generated by the KN mechanism.
We assume that this is the case.
Using (\ref{saxion-decay-width}), one then finds that $B_a$ has a value between
${\cal O}(10^{-4})$ and ${\cal O}(10^{-2})$ for $m_\sigma>2M_Z$.
Depending on $m_\sigma$, the decay is dominated by $\sigma\to hh$, $\sigma\to ZZ$,
or $\sigma\to t\bar t$ with $t$ being the top quark.
On the other hand, if $m_\sigma<2M_Z$, the saxion mainly decays into bottom quarks,
while giving ${\cal O}(10^{-3})\leq B_a<0.1$ for $m^2_h\leq 0.1|\mu|^2$.
Therefore, one can naturally avoid production of too many axions from the saxion
decay\footnote{
Models in \cite{thermal-inflation-axion,thermal-inflation-mirage-mediation} have
used this property to suppress the axion production from the saxion decay.
In the model of \cite{thermal-inflation-mirage-mediation} where the $\mu$ term
is generated by the GM mechanism, the authors introduced additional SM singlet
to achieve $C_\sigma={\cal O}(1)$.
On the other hand, in axionic mirage mediation \cite{axionic-mirage-mediation}
which also incorporates the GM mechanism, the large entropy released by the modulus
decay dilutes the axions produced by the saxion decay with $B_a\simeq 1$.
}.

There is also a constraint from the dark matter abundance.
First of all, the total abundance of LSPs should not exceed the measured
abundance of the dark matter\footnote{
The axion also contributes to the energy density of dark matter,
$\Omega_a \sim 0.4 \theta^2_a (S_0/10^{12}{\rm GeV})^{1.18}$ with
$\theta_a$ being the initial misalignment angle.
}.
In addition, an attention should be paid to the LSPs produced
from late decays of the axino/gravitino because they will contribute to the energy
density of hot dark matter.
The current energy density of cold dark matter is $\Omega_{\rm CDM}\simeq 0.2$,
while the density of hot dark matter is bounded from above as
$\Omega_{\rm HDM}\lesssim 10^{-3}$ \cite{PDG,reionisation}.
In the situation under consideration, the dark matter abundance essentially depends
on the decay temperature of the saxion
\bea
\hspace{-0.5cm}
T_\sigma = \left(\frac{90}{\pi^2g_\ast(T_\sigma)}\right)^{1/4}
\sqrt{\Gamma_\sigma M_{Pl}}
\simeq
6{\rm GeV}
\left(\frac{0.01}{B_a}\right)^{1/2}
\left(\frac{m_\sigma}{10^2{\rm GeV}}\right)^{3/2}
\left(\frac{10^{11}{\rm GeV}}{S_0}\right),
\eea
for $g_\ast(T_\sigma)={\cal O}(10^2)$, with $g_\ast$ being the effective degrees
of freedom of the radiation.
Note that there are two main processes for the production of axinos and
gravitinos, $(i)$ the decay of the saxion $\sigma$, and $(ii)$ the decay of the
LOSP $\tilde\chi$.
If $m_\sigma>2m_{\tilde \chi}$, LOSPs will be produced directly from
the saxion decay.
In addition, there are thermally generated LOSPs when $T_\sigma$ is higher
than $T_f$.
Here $T_f\sim m_\chi/20$ denotes the freeze-out temperature
of $\tilde\chi$, below which $\tilde \chi$ decouples from the thermal bath.

We first concentrate on the direct production of axinos and gravitinos from
the saxion decay.
The yield of axinos produced from the saxion decay is the sum of the axino yield
from $\sigma\to \tilde a\tilde a$, $Y^\sigma_0$ and that from
$\sigma\to \tilde a\tilde G$, $Y^\sigma_{3/2}$:
\bea
Y^\sigma_{\tilde a} = Y^\sigma_0 + Y^\sigma_{3/2}.
\eea
Using the relation (\ref{saxion-to-LSPs}) and $m_{\tilde a}={\cal O}(M_0/8\pi^2)$,
each term is evaluated as
\bea
Y^\sigma_0 &=&
\frac{3}{2}\frac{T_\sigma}{m_\sigma}
\frac{\Gamma_{\sigma\to \tilde a \tilde a}}{\Gamma_\sigma}
\nonumber \\
&\sim&
4\times 10^{-10}
\left(\frac{c_{\tilde a}}{0.1}\right)^2
\left(\frac{B_a}{0.01}\right)^{1/2}
\left(\frac{m_{\tilde a}}{1{\rm GeV}}\right)^2
\left(\frac{10^2{\rm GeV}}{m_\sigma}\right)^{3/2}
\left(\frac{10^{11}{\rm GeV}}{S_0}\right),
\eea
and
\bea
\label{LSP-yield-saxion}
Y^\sigma_{3/2} = \frac{3}{4}\frac{T_\sigma}{m_\sigma}
\frac{\Gamma_{\sigma\to \tilde a \tilde G}}{\Gamma_\sigma}
\sim
10^{-4}\left(\frac{0.1}{c_{\tilde a}}\right)^2
\left(\frac{m_\sigma}{0.3 M_0}\right)^4
\left(\frac{S_0}{10^{-2}X}\right)^2Y^\sigma_0 \ll Y^\sigma_0,
\eea
the latter of which is the same as the gravitino yield from
$\sigma\to \tilde a\tilde G$.
Note that, when the saxion is stabilized with
$y^2_\Psi S^2_0\leq {\cal O}(10^{-3}\Lambda^2_\Phi)$, one generically obtains
$c_{\tilde a}={\cal O}(0.1)$ and $m_\sigma={\cal O}(0.1 M_0)$.
For $y_\Psi S_0={\cal O}(0.1\Lambda_\Phi)$, $c_{\tilde a}$ is of order
unity, but the saxion acquires mass of ${\cal O}(M_0)$.
As discussed already, $B_a<0.1$ can be achieved for $m^2_\sigma\ll |\mu|^2$.
To be consistent with the cosmological observations, $Y^\sigma_{\tilde a,3/2}$
should satisfy the constraints
\bea
\label{cos-constraint}
Y^\sigma_{\tilde a} \leq 3.6\times 10^{-10}
\left(\frac{1{\rm GeV}}{m_{\tilde a}}\right)
\,&{\rm and}&\,
Y^\sigma_{3/2} \lesssim 1.8\times 10^{-12}
\left(\frac{1{\rm GeV}}{m_{\tilde a}}\right),
\eea
for the axino LSP, where the latter one comes from the constraint on the hot
component.
For the gravitino LSP, the produced axinos decay into gravitinos, yielding the
hot dark matter.
Thus, we obtain the constraint
\bea
Y^\sigma_{\tilde a} \lesssim 1.8\times 10^{-12}
\left(\frac{1{\rm GeV}}{m_{3/2}}\right).
\eea
It is interesting to note that, if the axino is the LSP, the above cosmological
constraints can be satisfied easily for $S_0\gtrsim 10^{11}$GeV.
In addition, the axino can naturally explain the dark matter component
of the Universe today.
On the other hand, in the case of the gravitino LSP, the constraint is severer
but not difficult to satisfy.
For instance, models with $S_0\sim 10^{11}$GeV and $m_{3/2}\sim 10$MeV will
survive the constraint.
The axino yield is further suppressed when $B_a$ is less than $0.01$.
Here, one should note that the relation (\ref{LSP}) leads to
$(m_{\tilde a}/m_{3/2})S_0\lesssim 10^{13}$GeV for $S_0\leq 0.1X$.

Another process for the production of axino/gravitino is the LOSP decay.
This becomes important either when $T_\sigma>T_f$, or when $T_\sigma<T_f$
and $m_\sigma>2m_{\tilde \chi}$.
For $S_0<0.1X$, since $\tilde \chi$ will dominantly decay into axinos with
$\Gamma_{\tilde\chi\to\tilde G}\ll 10^{-3}\Gamma_{\tilde \chi\to\tilde a}$,
the gravitino production from its decay can safely be ignored.
Let us examine the case that $T_\sigma$ is above $T_f$, for which LOSPs
can be in thermal equilibrium with SM particles.
The axino production is then determined by the total decay width of $\tilde\chi$
\bea
\Gamma_{\tilde \chi} \equiv \frac{r^{\tilde \chi}_{\tilde a}}{4\pi}
\frac{m^3_{\tilde \chi}}{S^2_0}.
\eea
A stau LOSP has $r^{\tilde \chi}_{\tilde a}={\cal O}(10^{-3})$, whereas
a bino LOSP has ${\cal O}(10^{-4})\leq r^{\tilde \chi}_{\tilde a} \leq{\cal O}(10^{-2})$,
depending on their mass.
The decay rate $\Gamma_{\tilde \chi}$ is thus smaller than the Hubble parameter
at $T=T_\sigma$, for $S_0\gtrsim 10^{10}$GeV.
If the decay temperature is below $m_{\tilde \chi}$, the yield of axinos is naively
estimated as $Y^{\tilde \chi}_{\tilde a} = Y^{\tilde \chi}_0 +Y^{\tilde \chi}_1$,
with $Y^{\tilde\chi}_{0,1}$ given by
\bea
\hspace{-0.6cm}
Y^{\tilde \chi}_0 &\sim&
\frac{45}{2\pi^3\sqrt{2\pi}}
\frac{1}{g_\ast(T_f)}
\left(\frac{m_{\tilde \chi}}{T_f}\right)^{3/2}e^{-m_{\tilde\chi}/T_f},
\nonumber \\
\hspace{-0.6cm}
Y^{\tilde \chi}_1 &\sim&
\frac{45}{2\pi^3\sqrt{2\pi}}
\int^{t_f}_{t_\sigma}
dt \frac{\Gamma_{\tilde \chi}}{g_\ast(T)}
\left(\frac{m_{\tilde \chi}}{T}\right)^{3/2}e^{-m_{\tilde \chi}/T}
\nonumber \\
&\simeq&
3\times10^{-9}
\left[ \int^{x_\sigma}_{x_f}dx \left(\frac{100}{g_\ast(x)}\right)^{3/2}
\frac{e^{-1/x}}{x^{9/2}}\right]
\left(\frac{r^{\tilde \chi}_{\tilde a}}{10^{-3}}\right)
\left(\frac{m_{\tilde \chi}}{200{\rm GeV}}\right)
\left(\frac{10^{11}{\rm GeV}}{S_0}\right)^2,
\eea
for $x\equiv T/m_{\tilde\chi}$ with
$x_\sigma=T_\sigma/m_{\tilde\chi}$ and $x_f=T_f/m_{\tilde\chi}$.
After freeze-out of $\tilde\chi$, all the remaining LOSPs will
decay into axinos.
This contribution gives $Y^{\tilde \chi}_0$, which is less than
${\cal O}(10^{-13})$ for $T_f\leq m_{\tilde \chi}/20$.
Hence, a constraint is placed on $Y^{\tilde \chi}_1$.
Substituting $Y^\sigma_{\tilde a}$ by $Y^{\tilde \chi}_1$ in the equation
(\ref{cos-constraint}), one can find that the LOSP decay would not cause
cosmological problems for an axino LSP if $T_\sigma\lesssim m_{\tilde \chi}/10$,
or $S_0\gtrsim 10^{11}$GeV.
For $S_0\sim 10^{10}$GeV, it is still possible to achieve $T_\sigma$ below
$m_{\tilde \chi}/10$ when the saxion acquires mass,
$m_\sigma={\cal O}(0.1M_0)$.
On the other hand, if the gravitino is the LSP, one would need
$T_\sigma\lesssim m_{\tilde \chi}/15$ and $m_{3/2}\ll 1$GeV to suppress
the abundance of hot LSPs.
Notice that the case with $T_\sigma>m_{\tilde \chi}$ should obviously
be excluded\footnote{
Axinos can also be produced by thermal scattering \cite{Thermal-axino}
after saxion decay.
But the thermal production would be negligible when the saxion decay
temperature is much lower than the MSSM sparticle masses.
}.

Let's move on to the case that $T_\sigma$ is lower than $T_f$,
for which there are only a negligible number of LOSPs in thermal bath.
However, LOSPs will be produced abundantly from the saxion decay if the decay
process is kinematically allowed.
In this case, the annihilation process can be effective to reduce their number
density, depending on the decay width of the saxion into $\tilde \chi$:
\bea
\Gamma_{\sigma\to\tilde\chi} \equiv r^\sigma_{\tilde\chi} B_a\Gamma_\sigma.
\eea
From (\ref{saxion-decay-width}), one obtains
${\cal O}(10^{-3})\leq r^\sigma_{\tilde\chi}\leq {\cal O}(10^{-1})$ for
a bino LOSP, while
${\cal O}(10^{-6})\leq r^\sigma_{\tilde\chi}\leq {\cal O}(10^{-3})$
in the stau LOSP case.
The LOSPs produced by the saxion decay will annihilate with each other
if the interaction rate is much larger than $\Gamma_{\tilde\chi}$.
This condition translates into
\bea
\hspace{-0.5cm}
\langle \sigma_{\rm ann} v_{\rm rel} \rangle_{\tilde\chi} \gg
\frac{10^{-18}}{{\rm GeV}^2}
\left(\frac{0.01}{B_a}\right)
\left(\frac{r^{\tilde\chi}_{\tilde a}}{10^{-3}}\right)
\left(\frac{10^{-3}}{r^\sigma_{\tilde\chi}}\right)
\left(\frac{m_\sigma}{2m_{\tilde\chi}}\right)
\left(\frac{m_{\tilde\chi}}{25T_\sigma}\right)^4
\left(\frac{10^{11}{\rm GeV}}{S_0}\right)^2,
\eea
where $\langle\cdots\rangle$ represents
the thermal average of the annihilation cross section times the relative
velocity of $\tilde\chi$.
For a bino or stau LOSP, the above condition is indeed satisfied well,
implying that LOSPs are so abundant.
Therefore, the annihilation process occurs quite effectively until
the Hubble parameter becomes comparable to the annihilation rate.
After annihilation, $\tilde \chi$ decays into an axino and axion, and thus
the axino abundance is determined by
\bea
Y^{\tilde\chi\prime}_{\tilde a}
&\simeq& Y^\sigma_{\tilde\chi}
\simeq \frac{1}{4}
\left(\frac{90}{\pi^2g_\ast(T_\sigma)}\right)^{1/2}
\frac{1}{\langle \sigma_{\rm ann} v_{\rm rel}
\rangle_{\tilde\chi} T_\sigma M_{Pl}}
\nonumber \\
&\simeq&
3\times 10^{-12}
\left(\frac{100}{g_\ast(T_\sigma)}\right)^{1/2}
\left(\frac{10^{-8}{\rm GeV}^{-2}}{\langle \sigma_{\rm ann} v_{\rm rel}
\rangle_{\tilde\chi}}\right)
\left(\frac{1{\rm GeV}}{T_\sigma}\right).
\eea
For a stau LOSP, the annihilation cross section is roughly given by
$\langle \sigma_{\rm ann} v_{\rm rel} \rangle_{\tilde \tau}\sim
10 \alpha^2_{\rm em} m^{-2}_{\tilde \tau}$ \cite{stau-annihilation},
and has a value of ${\cal O}(10^{-8}){\rm GeV}^{-2}$
for $m_{\tilde \tau}=200$GeV.
A bino LOSP has $\langle \sigma_{\rm ann} v_{\rm rel} \rangle_{\tilde B}\sim
4 \alpha^2_1 m^2_t m^{-4}_{\tilde t_R}$ \cite{bino-annihilation}, which is of
${\cal O}(10^{-9}){\rm GeV}^{-2}$ for $m_{\tilde t_R}=500$GeV.
In addition, $T_\sigma$ cannot be much lower than $0.1$GeV
for $10^{10}{\rm GeV}\lesssim S_0\lesssim10^{12}{\rm GeV}$.
It is thus not difficult to make $Y^{\tilde\chi\prime}_{\tilde a}$ less than
$Y^\sigma_{\tilde a}$.
This implies that, even for $T_\sigma$ below $T_f$, the saxion can have
mass $m_\sigma>2m_{\tilde \chi}$ without causing cosmological difficulties.

We close this section by summarizing the implications of the Higgs
coupling to $S$.
As discussed already, the coupling between $H_uH_d$ and $S$ naturally
explains the $\mu$ and $B\mu$ terms in the MSSM.
Furthermore, it makes the saxion and axino interact with the MSSM particles
via various couplings suppressed by the PQ scale.
Mediated by these interactions, the LOSP decay into axinos can occur inside
the detector, depending on the PQ scale.
The decay length will give us a direct information on the PQ scale.
On the other hand, the saxion products might cause cosmological problems,
once the Universe is dominated by the saxion.
Since the saxion decays into axions, axinos, and gravitinos, there are
constraints from the Big Bang nucleosynthesis and the dark matter abundance.
However, the cosmological constraints can naturally be satisfied when the $\mu$
term is generated via the KN mechanism, for which the saxion coupling to
the MSSM particles is stronger than those induced by the PQ messenger
loops.

\section{Conclusion}

In this paper, we have studied a simple axion model that establishes a connection between
the origin of the Higgs $\mu/B\mu$ term and the solution to the strong CP
problem within the framework of gauge mediation.
Such a connection is possible if the model possesses the PQ symmetry
under which the Higgs bilinear $H_uH_d$ is charged.
The PQ symmetry breaking is governed by SUSY breaking effects.
We pointed out that a crucial role is played by the mixing between
the messengers transmitting the SUSY breaking and the PQ symmetry breaking
to the MSSM sector.
In the presence of such mixing, the PQ scale is radiatively stabilized
at a scale below the gauge mediation scale.
This stabilization mechanism can apply to other cases as well, such as
models with a generalized messenger sector, or flaton models where $S$
corresponds to a flaton field.

Also important is that the model provides a natural explanation for
the presence of both $\mu$ and $B\mu$.
They are generated with the correct size from a coupling between $H_uH_d$
and $S$, which also induces the saxion/axino interactions with the MSSM
particles.
Furthermore, the phase of $B$ is aligned with that of the gaugino masses,
thereby not spoiling the nice property of gauge mediation that the induced
soft terms do not lead to excessive flavor and CP violations.
In the model, the LSP is either the axino or the gravitino depending on
the scale of gauge mediation, while the LOSP can be the bino or the stau.
The Higgs coupling to $S$ leads to that the LOSP mainly decays into axinos
with coupling suppressed by the PQ scale.
Thus, the collider signature highly depends on the PQ scale.
If the saxion is stabilized at a scale around $10^{10}$GeV or less,
the LOSP can decay within the detector while giving distinct signals.
On the other hand, for $S_0$ larger than $10^{10}$GeV, the LOSP will decay with
a rather long lifetime, but still a non-negligible amount of the LOSPs will
decay inside the detector unless $S_0$ is out of the axion window.

We have also investigated the cosmological constraints placed on the saxion when
it dominates the energy density of the Universe.
In order for the saxion decay not to conflict with the successful predictions
of the Big Bang nucleosynthesis, its branching ratio into axions should be
suppressed.
Moreover, the LSPs from the saxion decay should not overclose the Universe.
In particular, the hot LSPs from the late decay of axino/gravitino should
be small enough to be consistent with the cosmological observations.
All these constraints can be satisfied well for the PQ scale
around $10^{11}$GeV when the $\mu$ term is generated via the KN mechanism,
i.e. from a superpotential Higgs coupling to $S$.
This is because the saxion is coupled to the SM particles more strongly compared
to the case of the GM mechanism.
If there is an extra entropy production, models that incorporate the GM mechanism
can still be cosmologically viable.

\acknowledgments

We thank Hyung Do Kim for pointing out the sign error in the saxion soft
mass in the previous version.
KSJ thanks Eung Jin Chun for discussion on the thermal production of axinos.
This work was supported by the JSPS Grant-in-Aid 21-09224 for JSPS Fellows.

\appendix

\section{PQ sector soft terms}

In this appendix, we provide the expressions for PQ sector soft terms
for the case that the saxion is stabilized at
$y_\Psi|S|\leq{\cal O}(\Lambda_\Phi/\sqrt{8\pi^2})$.
The soft terms are parameterized in terms of
$\{M_0,\Lambda_\Phi,N_\Phi,\Lambda_\Psi,N_\Psi\}$, and the supersymmetric
couplings $g_a$ and $y_{q,\ell}$.
The Yukawa couplings are written as
\bea
y_\Psi S \Psi\bar\Psi= y_q S q\bar q + y_\ell S \ell\bar\ell,
\eea
for the PQ triplet $q$ and doublet $\ell$.
Integrating out $\Phi+\bar\Phi$, soft terms are generated as
\bea
\label{PQ-soft-terms}
\frac{A_{q,\ell}(\Lambda^-_\Phi)}{M_0} &=& N_\Phi\left(
10N_\Psi + 1 + 16\pi^2\frac{|S|^2}{\Lambda^2_\Phi}\right)y^2_S,
\nonumber \\
\frac{m^2_{\tilde{\bar q}}(\Lambda^-_\Phi)}{M^2_0} &=&
N_\Phi \left(
\frac{8}{3}g^4_3 + \frac{2}{15}g^4_1 - (5N_\Psi+1)y^2_\Psi y^2_S
\right),
\nonumber \\
\frac{m^2_{\tilde{\bar \ell}}(\Lambda^-_\Phi)}{M^2_0} &=&
N_\Phi \left(
\frac{3}{2}g^4_2 + \frac{3}{10}g^4_1 - (5N_\Psi+1)y^2_\Psi y^2_S
\right),
\nonumber \\
\frac{m^2_{\tilde q}(\Lambda^-_\Phi)}{M^2_0} &=&
N_\Phi \left(
\frac{8}{3}(g^2_3-2y^2_S)g^2_3
+ \frac{2}{15}(g^2_1-2y^2_S)g^2_1
+ \xi y^2_S \right),
\nonumber \\
\frac{m^2_{\tilde \ell}(\Lambda^-_\Phi)}{M^2_0} &=&
N_\Phi \left(
\frac{3}{2}(g^2_2-2y^2_S)g^2_2 + \frac{3}{10}(g^2_1-2y^2_S)g^2_1
+ \xi y^2_S \right),
\eea
at the scale just below $\Lambda_\Phi$, with
\bea
\xi = (5 N_\Phi N_\Psi+N_\Phi+N_\Psi)y^2_S + y^2_\Phi
- (16\pi^2)^2 \frac{|S|^2}{\Lambda^2_\Phi},
\eea
where the gauge and Yukawa couplings are evaluated at $\Lambda_\Phi$,
neglecting the splitting in PQ Yukawa couplings.

At scales between $\Lambda_\Psi$ and $\Lambda_\Phi$, soft terms are RG
evolved as
\bea
\frac{d A_q}{d\ln Q} &=& \frac{1}{8\pi^2}\left\{
(3N_\Psi+2)y^2_q A_q + 2N_\Psi y^2_\ell A_\ell
-2 \left( \frac{8}{3}g^2_3M_3 + \frac{2}{15}g^2_1 M_1
\right)\right\},
\nonumber \\
\frac{d A_\ell}{d\ln Q} &=&
\frac{1}{8\pi^2}\left\{
3N_\Psi y^2_q A_q + (2N_\Psi+2) y^2_\ell A_\ell
-2 \left( \frac{3}{2}g^2_2M_2 + \frac{3}{10}g^2_1 M_1
\right)\right\},
\nonumber \\
\frac{d m^2_{\tilde q,\tilde {\bar q}}}{d\ln Q} &=&
\frac{1}{8\pi^2}\left\{
y^2_q P_{\tilde q}
-2 \left( \frac{8}{3}g^2_3|M_3|^2
+ \frac{2}{15}g^2_1|M_1|^2 \right) \right\},
\nonumber \\
\frac{d m^2_{\tilde \ell,\tilde {\bar \ell}}}{d \ln Q} &=&
\frac{1}{8\pi^2}\left\{
y^2_\ell P_{\tilde \ell}
-2 \left( \frac{3}{2}g^2_2|M_2|^2
+ \frac{3}{10}g^2_1|M_1|^2 \right) \right\},
\nonumber \\
\frac{d m^2_S}{d\ln Q} &=&
\frac{N_\Psi}{8\pi^2} \left(
3y^2_q P_{\tilde q} + 2y^2_\ell P_{\tilde \ell} \right),
\eea
for $P_{\tilde q} = m^2_S +m^2_{\tilde q} + m^2_{\tilde{\bar q}}+|A_q|^2$,
and $P_{\tilde \ell} = m^2_S +m^2_{\tilde \ell} + m^2_{\tilde{\bar \ell}}+|A_\ell|^2$.
On the other hand, the running of PQ Yukawa couplings is determined by
\bea
\frac{d y^2_q}{d\ln Q} &=& \frac{y^2_q}{8\pi^2}\left\{
(3N_\Psi+2)y^2_q + 2N_\Psi y^2_\ell
-2\left( \frac{8}{3}g^2_3 + \frac{2}{15}g^2_1 \right)\right\},
\nonumber \\
\frac{d y^2_\ell}{d\ln Q} &=& \frac{y^2_\ell}{8\pi^2}\left\{
3N_\Psi y^2_q + (2N_\Psi+2)y^2_\ell
-2\left( \frac{3}{2}g^2_2 + \frac{3}{10}g^2_1 \right)\right\},
\eea
with gauge couplings given by
\bea
\frac{1}{g^2_a(Q)} &\simeq& 2
+ \frac{N_\Psi}{8\pi^2}\ln\left(\frac{\Lambda_\Psi}{Q}\right)
+ \frac{b_a}{8\pi^2}\ln\left(\frac{M_{\rm GUT}}{Q}\right),
\eea
at $\Lambda_\Psi<Q<\Lambda_\Phi$.
Here $b_a$ are the beta function coefficients for the MSSM.

\section{Mixing parameters}

To derive the saxion/axino couplings to MSSM particles, one can make
the replacements (\ref{replacement-saxion}) and (\ref{replacement-axino}).
The mixing between the saxion with neutral CP even Higgs bosons, $h$ and $H$,
is parameterized by $N^\sigma_{d,u}$:
\bea
(N^\sigma_d,N^\sigma_u) =
(-n_\sigma \sin\alpha + n^\prime_\sigma \cos\alpha,
n_\sigma \cos\alpha+n^\prime_\sigma \sin\alpha),
\eea
where $n_\sigma$ and $n^\prime_\sigma$ are given by
\bea
n_\sigma &=& 2\sin(\beta-\alpha)-\frac{|B|}{|\mu|}\cos(\beta+\alpha),
\nonumber \\
n^\prime_\sigma &=&
\frac{m^2_h-m^2_\sigma}{m^2_H-m^2_\sigma}\left(
2\cos(\beta-\alpha)-\frac{|B|}{|\mu|}\sin(\beta+\alpha) \right),
\eea
with $\alpha$ being the mixing angle for $h$ and $H$.
On the other hand, the parameters $N^{\tilde a}_i$ for the axino mixing
with the neutral gauginos and Higgsinos are determined by
\bea
(N^{\tilde a}_{\tilde B},N^{\tilde a}_{\tilde W^0}) &\simeq&
\frac{\cos2\beta}{1-n_{\tilde a}\sin2\beta}\left(
\frac{M_Z}{M_{\tilde B}}\sin\theta_W,
\frac{M_Z}{M_{\tilde W}}\cos\theta_W \right),
\nonumber \\
(N^{\tilde a}_{\tilde H^0_d},N^{\tilde a}_{\tilde H^0_u}) &\simeq&
\frac{1}{1-n_{\tilde a}\sin2\beta} (
\cos\beta-n_{\tilde a} \sin\beta, \sin\beta-n_{\tilde a} \cos\beta),
\eea
for $n_{\tilde a}$ defined by
\bea
n_{\tilde a} = \frac{M_Z}{\mu} \left( \frac{M_Z}{M_{\tilde B}} \sin^2\theta_W
+ \frac{M_Z}{M_{\tilde W}} \cos^2\theta_W \right),
\eea
where $M_Z$ is the Z boson mass, and $\theta_W$ is the weak mixing
angle.
Here we have neglected corrections suppressed by
$m_{\tilde a}/M_{\tilde B,\tilde W}$ or by $m_{\tilde a}/\mu$,
and have used that there is mixing between the neutral Higgsinos
and gauginos
\bea
{\cal L}^{\tilde H}_{\rm mix} = M_Z \left(
\tilde H^0_d \cos\beta - \tilde H^0_u \sin\beta \right)
\left(\tilde B\sin\theta_W - \tilde W^0\cos\theta_W \right)
+ {\rm h.c.},
\eea
which arises after the electroweak symmetry breaking.


\begin{thebibliography}{99}


\bibitem{gauge-mediation-classic}
  M.~Dine, W.~Fischler and M.~Srednicki,
  ``Supersymmetric Technicolor,''
  Nucl.\ Phys.\  B {\bf 189}, 575 (1981);
  S.~Dimopoulos and S.~Raby,
  ``Supercolor,''
  Nucl.\ Phys.\  B {\bf 192}, 353 (1981);
  M.~Dine and W.~Fischler,
  ``A Phenomenological Model Of Particle Physics Based On Supersymmetry,''
  Phys.\ Lett.\  B {\bf 110}, 227 (1982);
  C.~R.~Nappi and B.~A.~Ovrut,
  ``Supersymmetric Extension Of The SU(3) X SU(2) X U(1) Model,''
  Phys.\ Lett.\  B {\bf 113}, 175 (1982);
  L.~Alvarez-Gaume, M.~Claudson and M.~B.~Wise,
  ``Low-Energy Supersymmetry,''
  Nucl.\ Phys.\  B {\bf 207}, 96 (1982);
  S.~Dimopoulos and S.~Raby,
  ``Geometric Hierarchy,''
  Nucl.\ Phys.\  B {\bf 219}, 479 (1983).


\bibitem{gauge-mediation}
  M.~Dine and A.~E.~Nelson,
  ``Dynamical supersymmetry breaking at low-energies,''
  Phys.\ Rev.\  D {\bf 48}, 1277 (1993)
  [arXiv:hep-ph/9303230];
  M.~Dine, A.~E.~Nelson and Y.~Shirman,
  ``Low-Energy Dynamical Supersymmetry Breaking Simplified,''
  Phys.\ Rev.\  D {\bf 51}, 1362 (1995)
  [arXiv:hep-ph/9408384];
  M.~Dine, A.~E.~Nelson, Y.~Nir and Y.~Shirman,
  ``New tools for low-energy dynamical supersymmetry breaking,''
  Phys.\ Rev.\  D {\bf 53}, 2658 (1996)
  [arXiv:hep-ph/9507378];
  G.~F.~Giudice and R.~Rattazzi,
  ``Theories with gauge-mediated supersymmetry breaking,''
  Phys.\ Rept.\  {\bf 322}, 419 (1999)
  [arXiv:hep-ph/9801271].


\bibitem{review-gauge-mediation}
  For a review, see
  R.~Kitano, H.~Ooguri, Y.~Ookouchi,
  ``Supersymmetry Breaking and Gauge Mediation,''
  [arXiv:1001.4535 [hep-th]].


\bibitem{PQ-mechanism}
  R.~D.~Peccei and H.~R.~Quinn,
  ``CP Conservation In The Presence Of Instantons,''
  Phys.\ Rev.\ Lett.\  {\bf 38}, 1440 (1977);
  ``Constraints Imposed By CP Conservation In The Presence Of Instantons,''
  Phys.\ Rev.\  D {\bf 16}, 1791 (1977).


\bibitem{review-axion}
  For a review, see
  J.~E.~Kim,
  ``Light Pseudoscalars, Particle Physics and Cosmology,''
  Phys.\ Rept.\  {\bf 150}, 1 (1987);
  H.~Y.~Cheng,
  ``The Strong CP Problem Revisited,''
  Phys.\ Rept.\  {\bf 158}, 1 (1988);
  J.~E.~Kim and G.~Carosi,
  ``Axions and the Strong CP Problem,''
  arXiv:0807.3125 [hep-ph].


\bibitem{Kim-Nilles-mechanism}
  J.~E.~Kim and H.~P.~Nilles,
  ``The Mu Problem And The Strong CP Problem,''
  Phys.\ Lett.\  B {\bf 138}, 150 (1984).


\bibitem{chun-mu-axion}
  E.~J.~Chun, J.~E.~Kim and H.~P.~Nilles,
  ``A Natural solution of the mu problem with a composite axion in the hidden
  sector,''
  Nucl.\ Phys.\  B {\bf 370}, 105 (1992).


\bibitem{murayama-mu-axion}
  H.~Murayama, H.~Suzuki and T.~Yanagida,
  ``Radiative breaking of Peccei-Quinn symmetry at the intermediate mass
  scale,''
  Phys.\ Lett.\  B {\bf 291}, 418 (1992).


\bibitem{martin-axion-mu-problem}
  S.~P.~Martin,
  ``Collider signals from slow decays in supersymmetric models with an
  intermediate-scale solution to the mu problem,''
  Phys.\ Rev.\  D {\bf 62}, 095008 (2000)
  [arXiv:hep-ph/0005116].


\bibitem{KSVZ}
  J.~E.~Kim,
  ``Weak Interaction Singlet And Strong CP Invariance,''
  Phys.\ Rev.\ Lett.\  {\bf 43}, 103 (1979);
  M.~A.~Shifman, A.~I.~Vainshtein and V.~I.~Zakharov,
  ``Can Confinement Ensure Natural CP Invariance Of Strong Interactions?,''
  Nucl.\ Phys.\  B {\bf 166}, 493 (1980).


\bibitem{DFSZ}
  A.~R.~Zhitnitsky,
  ``On Possible Suppression Of The Axion Hadron Interactions. (In Russian),''
  Sov.\ J.\ Nucl.\ Phys.\  {\bf 31} (1980) 260
  [Yad.\ Fiz.\  {\bf 31} (1980) 497];
  M.~Dine, W.~Fischler and M.~Srednicki,
  ``A Simple Solution To The Strong CP Problem With A Harmless Axion,''
  Phys.\ Lett.\  B {\bf 104}, 199 (1981).


\bibitem{saxion-run-away-gauge-mediation}
  N.~Arkani-Hamed, G.~F.~Giudice, M.~A.~Luty and R.~Rattazzi,
  ``Supersymmetry-breaking loops from analytic continuation into superspace,''
  Phys.\ Rev.\  D {\bf 58}, 115005 (1998)
  [arXiv:hep-ph/9803290].


\bibitem{hadronic-axion-gauge-mediation}
  T.~Asaka and M.~Yamaguchi,
  ``Hadronic axion model in gauge-mediated supersymmetry breaking,''
  Phys.\ Lett.\  B {\bf 437}, 51 (1998)
  [arXiv:hep-ph/9805449];
  ``Hadronic axion model in gauge-mediated supersymmetry breaking and
  cosmology of saxion,''
  Phys.\ Rev.\  D {\bf 59}, 125003 (1999)
  [arXiv:hep-ph/9811451].


\bibitem{Chun-axion-mu-gauge-mediation}
  E.~J.~Chun,
  ``Strong CP and mu problem in theories with gauge mediated supersymmetry breaking,''
  Phys.\ Rev.\  {\bf D59}, 015011 (1999).
  [hep-ph/9712406].


\bibitem{axions-in-GM}
  L.~M.~Carpenter, M.~Dine, G.~Festuccia and L.~Ubaldi,
  ``Axions in Gauge Mediation,''
  Phys.\ Rev.\  D {\bf 80}, 125023 (2009)
  [arXiv:0906.5015 [hep-th]].


\bibitem{thermal-inflation}
  D.~H.~Lyth and E.~D.~Stewart,
  ``Cosmology With A Tev Mass GUT Higgs,''
  Phys.\ Rev.\ Lett.\  {\bf 75}, 201 (1995)
  [arXiv:hep-ph/9502417];
  ``Thermal Inflation And The Moduli Problem,''
  Phys.\ Rev.\  D {\bf 53}, 1784 (1996)
  [arXiv:hep-ph/9510204].


\bibitem{Analytic-cont}
  G.~F.~Giudice and R.~Rattazzi,
  ``Extracting supersymmetry breaking effects from wave function
  renormalization,''
  Nucl.\ Phys.\  B {\bf 511}, 25 (1998)
  [arXiv:hep-ph/9706540].


\bibitem{Giudice-Masiero-mechanism}
  G.~F.~Giudice and A.~Masiero,
  ``A Natural Solution to the mu Problem in Supergravity Theories,''
  Phys.\ Lett.\  B {\bf 206} (1988) 480.


\bibitem{small-b-gauge-mediation}
  G.~R.~Dvali, G.~F.~Giudice and A.~Pomarol,
  ``The $\mu$-Problem in Theories with Gauge-Mediated Supersymmetry Breaking,''
  Nucl.\ Phys.\  B {\bf 478}, 31 (1996)
  [arXiv:hep-ph/9603238].


\bibitem{sweet-spot-susy}
  M.~Ibe, R.~Kitano,
  ``Sweet Spot Supersymmetry,''
  JHEP {\bf 0708}, 016 (2007).
  [arXiv:0705.3686 [hep-ph]].
  

\bibitem{B-in-GM}
  G.~F.~Giudice, H.~D.~Kim and R.~Rattazzi,
  ``Natural mu and B mu in gauge mediation,''
  Phys.\ Lett.\  B {\bf 660}, 545 (2008)
  [arXiv:0711.4448 [hep-ph]].


\bibitem{domain-wall}
  S.~Kasuya, M.~Kawasaki and T.~Yanagida,
  ``Domain wall problem of axion and isocurvature fluctuations in chaotic
  inflation models,''
  Phys.\ Lett.\  B {\bf 415}, 117 (1997)
  [arXiv:hep-ph/9709202].


\bibitem{non-standard-gauge-mediation}
  N.~Arkani-Hamed, J.~March-Russell and H.~Murayama,
  ``Building models of gauge-mediated supersymmetry breaking without a
  messenger sector,''
  Nucl.\ Phys.\  B {\bf 509}, 3 (1998)
  [arXiv:hep-ph/9701286];
  E.~Poppitz and S.~P.~Trivedi,
  ``Some remarks on gauge-mediated supersymmetry breaking,''
  Phys.\ Lett.\  B {\bf 401}, 38 (1997)
  [arXiv:hep-ph/9703246].


\bibitem{axino-gravitino-collider}
  A.~Brandenburg, L.~Covi, K.~Hamaguchi, L.~Roszkowski and F.~D.~Steffen,
  ``Signatures of axinos and gravitinos at colliders,''
  Phys.\ Lett.\  B {\bf 617}, 99 (2005)
  [arXiv:hep-ph/0501287].


\bibitem{effective-gravitino-interaction}
  P.~Fayet,
  ``Lower Limit on the Mass of a Light Gravitino from e+ e- Annihilation
  Experiments,''
  Phys.\ Lett.\  B {\bf 175}, 471 (1986).


\bibitem{axino-dark-matter}
  L.~Covi, J.~E.~Kim and L.~Roszkowski,
  ``Axinos as cold dark matter,''
  Phys.\ Rev.\ Lett.\  {\bf 82}, 4180 (1999)
  [arXiv:hep-ph/9905212];
  L.~Covi, H.~B.~Kim, J.~E.~Kim and L.~Roszkowski,
  ``Axinos as dark matter,''
  JHEP {\bf 0105}, 033 (2001)
  [arXiv:hep-ph/0101009].


\bibitem{axino-cold-dark-matter}
  E.~J.~Chun, H.~B.~Kim, D.~H.~Lyth,
  ``Cosmological constraints on a Peccei-Quinn flatino as the lightest
  supersymmetric particle,''
  Phys.\ Rev.\  {\bf D62}, 125001 (2000).
  [hep-ph/0008139];
  E.~J.~Chun, H.~B.~Kim, K.~Kohri and D.~H.~Lyth,
  ``Flaxino dark matter and stau decay,''
  JHEP {\bf 0803}, 061 (2008)
  [arXiv:0801.4108 [hep-ph]].


\bibitem{CMB}
  S.~Bashinsky and U.~Seljak,
  ``Signatures of relativistic neutrinos in CMB anisotropy and matter
  clustering,''
  Phys.\ Rev.\  D {\bf 69}, 083002 (2004)
  [arXiv:astro-ph/0310198].


\bibitem{hot-dark-matter}
  S.~Hannestad and G.~Raffelt,
  ``Cosmological mass limits on neutrinos, axions, and other light
  particles,''
  JCAP {\bf 0404}, 008 (2004)
  [arXiv:hep-ph/0312154];
  P.~Crotty, J.~Lesgourgues and S.~Pastor,
  ``Current cosmological bounds on neutrino masses and relativistic relics,''
  Phys.\ Rev.\  D {\bf 69}, 123007 (2004)
  [arXiv:hep-ph/0402049];
  S.~Hannestad, A.~Mirizzi, G.~G.~Raffelt and Y.~Y.~Y.~Wong,
  ``Cosmological constraints on neutrino plus axion hot dark matter: Update
  after WMAP-5,''
  JCAP {\bf 0804}, 019 (2008)
  [arXiv:0803.1585 [astro-ph]].


\bibitem{axions-from-saxion-decay}
  K.~Choi, E.~J.~Chun and J.~E.~Kim,
  ``Cosmological implications of radiatively generated axion scale,''
  Phys.\ Lett.\  B {\bf 403}, 209 (1997)
  [arXiv:hep-ph/9608222].


\bibitem{thermal-inflation-axion}
  S.~Kim, W.~I.~Park and E.~D.~Stewart,
  ``Thermal inflation, baryogenesis and axions,''
  JHEP {\bf 0901}, 015 (2009)
  [arXiv:0807.3607 [hep-ph]].


\bibitem{thermal-inflation-mirage-mediation}
  K.~Choi, K.~S.~Jeong, W.~I.~Park and C.~S.~Shin,
  ``Thermal inflation and baryogenesis in heavy gravitino scenario,''
  JCAP {\bf 0911}, 018 (2009)
  [arXiv:0908.2154 [hep-ph]].


\bibitem{axionic-mirage-mediation}
  S.~Nakamura, K.~i.~Okumura and M.~Yamaguchi,
  ``Axionic Mirage Mediation,''
  Phys.\ Rev.\  D {\bf 77}, 115027 (2008)
  [arXiv:0803.3725 [hep-ph]].


\bibitem{PDG}
  C.~Amsler {\it et al.}  [Particle Data Group],
  ``Review of particle physics,''
  Phys.\ Lett.\  B {\bf 667}, 1 (2008).


\bibitem{reionisation}
  K.~Jedamzik, M.~Lemoine and G.~Moultaka,
  ``Gravitino, Axino, Kaluza-Klein Graviton Warm and Mixed Dark Matter and
  Reionisation,''
  JCAP {\bf 0607}, 010 (2006)
  [arXiv:astro-ph/0508141].


\bibitem{Thermal-axino}
  K.~Rajagopal, M.~S.~Turner and F.~Wilczek,
  ``Cosmological implications of axinos,''
  Nucl.\ Phys.\  B {\bf 358}, 447 (1991);
  A.~Brandenburg and F.~D.~Steffen,
  ``Axino dark matter from thermal production,''
  JCAP {\bf 0408}, 008 (2004)
  [arXiv:hep-ph/0405158];
  A.~Strumia,
  ``Thermal production of axino Dark Matter,''
  JHEP {\bf 1006}, 036 (2010)
  [arXiv:1003.5847 [hep-ph]];
  E.~J.~Chun,
  ``Dark matter in the Kim-Nilles mechanism,''
  arXiv:1104.2219 [hep-ph].


\bibitem{stau-annihilation}
  T.~Asaka, K.~Hamaguchi and K.~Suzuki,
  ``Cosmological gravitino problem in gauge mediated supersymmetry breaking
  models,''
  Phys.\ Lett.\  B {\bf 490}, 136 (2000)
  [arXiv:hep-ph/0005136].


\bibitem{bino-annihilation}
  M.~Kawasaki, T.~Moroi and T.~Yanagida,
  ``Constraint on the Reheating Temperature from the Decay of the Polonyi
  Field,''
  Phys.\ Lett.\  B {\bf 370}, 52 (1996)
  [arXiv:hep-ph/9509399].


\end{thebibliography}
\end{document}